\documentclass[12pt]{article}
\usepackage[utf8]{inputenc}
\usepackage[russian,english]{babel} 
\usepackage[T2A]{fontenc}   
\usepackage{mathtools}
\usepackage{amsfonts,amssymb,amsmath} 
\usepackage{graphicx} 
\usepackage{cite}
\usepackage{hyperref}
\hypersetup{colorlinks=true, linkcolor=blue, filecolor=blue, urlcolor=blue} 

\newtheorem{proposition}{Proposition}

\allowdisplaybreaks

\newcommand{\om}{\omega}
\newcommand{\bk}{{\bf k}}
\newcommand{\D}{{\mathcal D}}
\newcommand{\Li}{{\mathcal L}}
\newcommand{\der}[2]{\dfrac{\mathrm{d} #1}{\mathrm{d} #2}}

\begin{document}

\title{\Large Control of the von Neumann Entropy
for an Open Two-Qubit System
Using Coherent and Incoherent Drives}

\author{ Oleg~V.~Morzhin\footnote{E-mail: \url{morzhin.oleg@yandex.ru};~ 
    \href{http://www.mathnet.ru/eng/person30382}{mathnet.ru/eng/person30382};~ 
    \href{https://orcid.org/0000-0002-9890-1303}{ORCID 0000-0002-9890-1303}} 
    \quad and \quad 
Alexander~N.~Pechen\footnote{E-mail: \url{apechen@gmail.com};~ 
    \href{http://www.mathnet.ru/eng/person17991}{mathnet.ru/eng/person17991};~ 
    \href{https://orcid.org/0000-0001-8290-8300}{ORCID 0000-0001-8290-8300}} 
    \vspace{0.2cm} \\
Department of Mathematical Methods for Quantum Technologies \& \\
Steklov International Mathematical Center, \\ 
Steklov Mathematical Institute of Russian Academy of Sciences, \\
8~Gubkina Str., 119991, Moscow, Russia} 

\date{}
\maketitle

\begin{abstract}
This article is devoted to developing an approach for manipulating  
the von Neumann entropy $S(\rho(t))$ of an open two-qubit system 
with coherent control and incoherent control inducing time-dependent decoherence rates. 
The following goals are considered: 
(a)~minimizing or maximizing the final entropy $S(\rho(T))$; 
(b)~steering $S(\rho(T))$ to a~given target value; 
(c)~steering $S(\rho(T))$ to a~target value and satisfying the pointwise state constraint $S(\rho(t)) \leq \overline{S}$ 
for a~given~$\overline{S}$; 
(d)~keeping $S(\rho(t))$ constant at a~given time interval. 
Under the Markovian dynamics determined by a Gorini--Kossakowski--Sudarshan--Lindblad type master equation, which contains coherent and incoherent controls, one- and two-step gradient projection methods and genetic algorithm have been adapted, taking into account the specifics of the objective functionals. 
The corresponding numerical results are provided and discussed.

\vspace{0.3cm}

\end{abstract} 

\section{Introduction}

The theory of (optimal) control of quantum systems 
(atoms, molecules, etc.) is important for developing quantum 
technologies~\cite{DongPetersenBook2023, KuprovBook2023, KochEPJQuantumTechnol2022,
AlessandroBook2021, KurizkiKofmanBook2021, KwonTomonagaEtAl2021,
BaiChenWuAn2021, Acin2018, KochJPhysCondensMatter2016,
DongWuYuanLiTarn2015, CongBook2014,
AltafiniTicozzi2012, BonnardBook2012, Gough_2012, Shapiro_Brumer_Book_2ndEd_2012,
BrifNewJPhys2010, FradkovBook2007, LetokhovBook2007, 
TannorBook2007, ButkovskyBook1990}. Modeling of control problems for  
quantum systems is based on various quantum mechanical equations with Markovian or non-Markovian dynamics, e.g.,~the~Schr{\"o}dinger, von~Neumann, Gorini--Kossakowski--Sudarshan--Lindblad (GKSL) equations, and~various objective functionals to be minimized or maximized. In~practical applications, often~the controlled quantum system is open, i.e.,~interacting with its environment, and this environment is considered as an~obstacle for controlling the system. However, in~some cases, one can use the environment as a~useful control resource, such as, for~example, in~the {\it incoherent control} 
approach~\cite{PechenRabitzPRA2006, PechenPRA2011}, 
where the spectral, generally time-dependent and non-equilibrium density of incoherent photons is used as a~control function jointly with the {\it coherent control} via lasers to manipulate such a~quantum system dynamics. Following this approach, various types and aspects of optimal control problems for one- and two-qubit systems were analyzed~\cite{MorzhinPechenQIP2023,
MorzhinPechenBulletinIrkutsk2023,
PetruhanovPechen2022, 
PetruhanovPechenJPA2023,
Morzhin_Pechen_IzvRAN2023,
Morzhin_Pechen_IJTP_2021}.  

One particularly important class of quantum control objectives includes thermodynamic quantities and entropy of the quantum system. Properties of the von Neumann entropy in general are discussed, e.g.,~in~\cite{Holevo_Book_2019, Wilde_Book_2017, Shirokov2010, NielsenChuangBook2010,Petz2001}. The~von Neumann entropy appears in various applied aspects of quantum theory, has applications in quantum communication and statistical physics~\cite{OhyaPetzBook,OhyaVolovichBook,Ohya2010,Bracken2020}, or~even in cross-linguistic comparisons of language networks~\cite{Vera2021}. The~von Neumann entropy of reduced density matrices of a~bipartite quantum system provides a~good measure of entanglement. It appears in various thermodynamic quantities, such as Helmholtz free energy, can serve as a~degree of purity of a~quantum state, etc. The system--bath interaction can play a~crucial role in  the emergence of the laws of thermodynamics from quantum consideration~\cite{Kosloff2013}. The control of dissipative quantum systems, which changes entropy of the quantum state, has been studied in various works. In~particular, an~analytical solution for the optimal control of a quantum dissipative three-level system leading to the decrease in entropy was provided~\cite{Sklarz_Tannor_Khaneja_2004}. Entropy production for controlled Markovian evolution was studied in~\cite{Pavon_Ticozzi_2007}. The~von Neumann entropy and R\'{e}nyi entropy changes for the laser cooling of molecules were investigated~\cite{BartanaKosloffTannor2001}. A~detailed study of entropy changing control targets is explored in~\cite{Kallush_Dann_Kosloff_2022}, when the external drive influences not only the primary system but also the dissipation induced by the environment. Similar to the control of entropy is the state-to-state control between two Gibbs states, which is used to accelerate thermalization and cool for an open system~\cite{Dann2020}. The~effects of the population decay, leading to the reduction of entropy, in~a~two-level Markovian dissipative system were considered in~\cite{Ohtsuki2023}. Reference~\cite{CanevaPRA2011} considers entanglement entropy maximization for the Lipkin--Meshkov--Glick model operating with $N=50$ spins and the subsystem with $L = N/2$ spins using the free-gradient chopped random basis (CRAB) ansatz. Non-Markovian regimes can also be effective, e.g.,~for quantum battery and heat machines~\cite{Uzdin2016}. Reference~\cite{AbeSasakiHaraTsumuraIFAC2008} considers a~stochastic master equation with a~finite-dimensional measurement-based quantum feedback control and linear entropy. Reference~\cite{SahraiArzhangSeifooryNavaeipour2013} considers an~open four-level atomic system and analyzes coherent control for the von Neumann entropy (total and reduced versions) via quantum interference. In~\cite{XingWuSciWorldJ2013, XingWuWCECS2013, XingHuangZhao2016}, a controller design approach for a~closed quantum system described by the Scr\"{o}dinger equation in terms of the von Neumann/Shannon entropy was proposed. Reference~\cite{AbbasKhudhairEtAl2022} considers the spatial control of entropy for a~three-level ladder-type atomic system that interacts with optical laser fields and an~incoherent pumping~field.

Reference \cite{PechenRabitzEPL2010} provides the formulation and analysis of control objectives describing optimization of thermodynamic quantities of the form $\langle O\rangle -\beta^{-1} S(\rho(T))$, where $O$ is some quantum observable (e.g., energy with Hamiltonian $H$), $\beta$ is inverse temperature, and~$S(\rho(T))$ is some concave type of entropy, e.g.,~the von Neumann entropy, of an~open quantum system density matrix at the final time~$T$. The~system evolution was considered as driven by some coherent and/or incoherent controls, including Markovian and non-Markovian cases, particularly  the cases of master equations with coherent and incoherent controls~\cite{PechenRabitzPRA2006}. The~objective was expressed as a~Mayer-type functional determined by the final state $\rho(T)$ ($\hat{\rho}_f$ at the final time~$t_f$ in the notations of~\cite{PechenRabitzEPL2010}). The~applied control $c=(u,n)$ (note that in~\cite{PechenRabitzEPL2010} this most general combination of coherent and incoherent controls was denoted by symbol $u$, which in the present work denotes only coherent control, whereas the combination of controls here we denote by $c$) directs the evolution of the system from the initial state to the final state and specifies the value of the objective which depends, through $\rho(T)$, on~the control~$c$. A~specific important example of such an objective is Helmholtz free energy, which corresponds to $O=H$. In~the case of trivial observable $O={\rm const}\cdot \mathbb I$, the~objective is reduced to the entropic form and differs from the entropy by a~non-essential for the optimization constant term. Based on this objective and Reference~\cite{PechenRabitzEPL2010}, we define below several other control problems involving~entropy.

The entropy of a quantum state was introduced by L. Landau to describe states of composite quantum systems~\cite{Landau1927}, which is related to using entropy as a measure of entanglement, and~by J.~von Neumann to describe the thermodynamic properties of quantum systems~\cite{vonNeumannBook}. This provides the motivation to introduce control problems focused on steering and maintaining the von Neumann entropy of system states. Objectives of forms~(\ref{J1_inf})--(\ref{J5_inf}) serve as examples of naturally extending problems related to maximizing or minimizing quantities involving entropy to controlling their behavior over a certain time range. Such a natural extension, in general, can include (but~is not limited to) the following: 
\begin{itemize}
\item Control the behavior of thermodynamic quantities, such as Helmholtz  free energy, not only at the final time instant but over some time range;
\item Control of the degree of entanglement of a bipartite system over time;
\item To not only maximize or minimize but~rather control the rate of entropy production.
\end{itemize} 

The basic task for all these problems is to manipulate entropy over a given time range which, including optimization methods, we consider in this work.

In general, quantum (open-loop) control, both for closed and open quantum systems, various types of optimization tools are~used: 
\begin{itemize}
\item For infinite-dimensional optimization, e.g.,
the Pontryagin maximum principle (PMP)~\cite{ButkovskyBook1990, BoscainPRXQuantum2021, BuldaevMathematics2022}, 
Krotov-type methods~(\cite{MorzhinPechenBulletinIrkutsk2023, Goerz_NJP_2014_2021, KrotovMorzhinTrushkova2013}, \cite{TannorBook2007}, \S~16.2.2, \cite{KrotovBook1996}, pp.~253--259), one- and two-step gradient projection methods (GPM-1, GPM-2)~\cite{Morzhin_Pechen_IJTP_2021, MorzhinPechenBulletinIrkutsk2023, MorzhinPechenQIP2023},~etc.;
\item For finite-dimensional optimization under various classes of parameterized controls, e.g.,~gradient ascent pulse engineering GRAPE-type methods (e.g., \cite{PetruhanovPechenJPA2023, PetruhanovPechen2022, Khaneja_JMagnReson_2005}, \cite[Section~3]{Morzhin_Pechen_IzvRAN2023}) (GRAPE-type methods operate with piecewise-constant controls, matrix exponentials, and~gradients), CRAB ansatz~\cite{CanevaPRA2011, MullerSaidJelezkoCalarcoMontangero2022} (coherent control is considered in terms of sine, cosine, etc.), genetic algorithm (GA)~\cite{JudsonRabitz1992, PechenRabitzPRA2006, BrownPaternostroFerraro2023}, dual annealing~\cite{MorzhinPechenBulletinIrkutsk2023},~etc.
\end{itemize}

In this article, we develop an approach for (open loop) control of the von Neumann entropy for open quantum systems driven by simultaneous coherent and incoherent controls. For~such~a system, we study control objectives based on the von Neumann entropy of the system states:
\begin{align}
S(\rho(t)) = - \mathrm{Tr} \left(\rho(t) \log \rho(t) \right) = -\sum_{\lambda_i(t) \neq 0} \lambda_i(t) \log \lambda_i(t),
\label{von_Neumann_entropy_rho_t}
\end{align}
where $\log$ denotes the natural matrix logarithm and $\lambda_i(t)$ are eigenvalues of $\rho(t)$.
For the initial time $t=0$ and final time $t=T$, we consider, correspondingly, 
$S(\rho_0)$ and $S(\rho(T))$. The~approach is based on using bounded coherent and incoherent controls to manipulate the von~Neumann entropy.  
Since the control of the entropy requires, in general, changing the degree of purity of the system density matrix, it requires the ability to generate a~given non-unitary dynamics. For~this, the combination of coherent and incoherent controls introduced in~\cite{PechenRabitzPRA2006} makes a~suitable~tool. 

To achieve these goals, we formulate the corresponding objective functionals. These functionals contain either differentiable or non-differentiable forms. For~the differentiable cases, both for the objective functionals of the Mayer and Mayer--Bolza types, we develop the 
one- and two-step GPMs for piecewise continuous controls based on deriving 
gradients of the objective functionals and the corresponding adjoint systems. For~the non-differentiable cases, piecewise linear controls are considered instead, and finite-dimensional optimization is performed using~GA. Moreover, various forms of regularization in controls are~provided.

The structure of the article is the following. In~Section~\ref{section2}, we briefly outline the incoherent control approach. In~Section~\ref{sec:objectives}, the objective functionals involving entropy for the described above problems are defined. In~Section~\ref{section4}, we consider---as an example---an open two-qubit system whose dynamics are determined by a GKSL-type master equation, which contains coherent and incoherent controls. Section~\ref{section5} describes the optimization approaches. Section~\ref{section6} provides and discusses the analytical and numerical results. Conclusions Section~\ref{Section_Conclusion} resumes the~article.

\section{Incoherent Control and Time-Dependent Decoherence~Rates} 
\label{section2}

The idea of incoherent control is to consider the environment as a useful resource for manipulating quantum systems. There are various approaches to using the environment as a~control. We exploit the idea proposed and developed for generic quantum systems in~\cite{PechenRabitzPRA2006, PechenPRA2011}. In~this approach, the state of the environment is used as a~control. Usually, the state of the environment is considered as the Gibbs (thermal) state with some temperature. However, the state of the environment can be a~more general non-thermal non-equilibrium state. If~the environment consists of photons, which is one of the most typical physical examples of the environment, its more general non-equilibrium state at some time instant $t$ is characterized by the distribution $n_{\bk,\alpha}(t)$ of photons in momenta 
$\bk$ and polarization $\alpha$. Moreover, this state and, hence, this distribution can evolve with time. Non-thermal distributions for photons are relatively easy to generate, so that it is a~physical and technically possible way of control. In~this work, we neglect polarization and directional dependence so that, here, the control is the distribution of photons only in frequency $\omega$ and time, $n_\om(t)$. In~the most general consideration, polarization and directional selectivity can be taken into account for the~control.

A time-evolving distribution of photons induced generally time-dependent decoherence rates of the system, which is immersed in this photonic environment, so that under certain approximations, the master equation for the system density matrix can be considered~as
\begin{equation}
\label{Eq:ME2} 
\der{\rho(t)}{t} = \Li^{u, n}_t (\rho(t)) := -i [H^{u, n}_t, \rho(t)] + \varepsilon\underbrace{\sum_k\gamma_k(t) \D_k (\rho(t))}_{\displaystyle\D^n_t (\rho(t))},\quad \rho(0) = \rho_0,\quad t\in[0,T].
\end{equation}
Here, both Markovian and non-Markovian cases can be included. The general formulation below is performed for both Markovian and non-Markovian cases, while only the Markovian case is explicitly analyzed. In~\cite{PechenRabitzPRA2006}, the~dissipators $\D_k$ corresponding to the weak coupling and low-density limits in the theory of open quantum systems were explicitly considered. In~general, other regimes, e.g., the ultrastrong coupling and the strong-decoherence limits~\cite{Trushechkin_2021,Trushechkin_2022}, or weakly damped quantum systems in various regimes~\cite{McCauley2020}, can be considered as well. For~the weak coupling limit case, the~decoherence rate for the transition between system states $|i\rangle$ and $|j\rangle$ with transition frequency $\omega_{ij}=E_{j}-E_{i}$ (here, $E_{i}$ is the energy of the system state $|i\rangle$) were considered in~\cite{PechenRabitzPRA2006} as
\begin{equation*}
\gamma_{ij}(t) = \pi\int  \,\delta(\omega_{ij}-\omega_{\bf k})|g({\bf k})|^2(n_{\omega_{ij}}(t)+\kappa_{ij})\mathrm{d} {\bf k},\qquad i,j = 1, \ldots, N.
\end{equation*}
Here, $\kappa_{ij}=1$ for $i>j$ and $\kappa_{ij}=0$; otherwise, $\omega_{\bf k}$ is the dispersion law for the bath (e.g., $\omega=|{\bf k}|c$ for photons, where ${\bf k}$ denotes photon momentum, $c$ denotes the speed of light), and~$g({\bf k})$ describes the coupling of the system to the $\bf k$-th mode of the photonic reservoir. For~$i>j$, the~summand $\kappa_{ij}=1$ describes spontaneous emission and $\gamma_{ij}$ determines the rate of both spontaneous and induced emissions between levels $i$ and $j$. For~$i<j$, $\gamma_{ij}$ determines the rate of induced absorption. These decoherence rates appear in~(\ref{Eq:ME2}), where $k=(i,j)$ is~multi-indexed. 

Such incoherent control appears to be rich enough to approximately generate, when combined with fast coherent control, arbitrary density matrices of generic quantum systems within the scheme proposed in~\cite{PechenPRA2011}. Hence, it can approximately realize the strongest possible degree of quantum state control---controllability of open quantum systems in the set of all density matrices. This scheme has several important features. (1)~It was obtained for a~physical class of dissipators $\D_k$ known in the weak coupling limit. (2)~It was obtained for generic quantum systems of an arbitrary dimension and for almost all values of the system parameters. (3)~A~simple explicit analytic solution for incoherent control was obtained. (4)~The control scheme is robust to variations of the initial state---the optimal control steers simultaneously {\it all} initial states into the target state, thereby physically realizing all-to-one Kraus maps theoretically exploited for quantum control in~\cite{Wu_2007_5681} and recently experimentally for an open single qubit in~\cite{Zhang_Saripalli_Leamer_Glasser_Bondar_2022}. In~\cite{PechenPRA2011}, coherent and incoherent controls were separated in time (first coherent control, followed by incoherent) and were applied to the system on different time scales determined by the parameters of the system. Incoherent control was applied on a time scale slower than coherent control. When coherent and incoherent controls are applied simultaneously, such a difference in time scales may lead to bounds on variations of incoherent control, considering that incoherent control should be varied slowly compared to coherent control. In~the analysis below, Equation~(\ref{obj_functions_with_regularization2}) is used to take into account such bounds on variations of the incoherent control. To shorten the incoherent control time scale, the first stage of the incoherent control scheme proposed in~{\cite{PechenPRA2011}} was further modified for a two-level system in~\cite{Morzhin_Pechen_IzvRAN2023}, significantly reducing the control time scale. Such an incoherent control can be technically implemented, e.g., as~it was done for controlling multi-species atomic and molecular systems with $\rm Gd_2O_2S\!:\!Er^{3+}$ (6\%) samples~\cite{LaforgeJCP2018}. 

\section{Control Objective Functionals Involving~Entropy}\label{sec:objectives}

In this section, we define control objective functionals, describing various problems involving entropy including both Markovian and non-Markovian cases.

Fixing $T,~\rho_0$, control $c = (u,n)$, $\varepsilon$, and~so on, one solves the initial problem~(\ref{Eq:ME2}) with the initial condition $\rho(0) = \rho_0$ to find the corresponding solution~$\rho$, a~matrix function defined at~$[0, T]$. For~each state $\rho(t)$, consider its von Neumann entropy $S(\rho(t))$. Using this standard notion of the von Neumann entropy, we formulate below several objective functionals based on the following objective functional for minimizing or maximizing the von Neumann entropy as considered in~\cite{PechenRabitzEPL2010}.
\begin{itemize}
\item {\it Minimizing or maximizing} the von Neumann entropy, or~more general thermodynamic quantities ($O$ is a~Hermitian observable, for~example, the Hamiltonian of the system, in~this case, it is Helmholtz free energy) at a~final time, as~defined in~\cite{PechenRabitzEPL2010}:
\begin{align}
	J_O(c)=\langle O\rangle -\frac{1}{\beta}S(\rho(T)) \to \inf/\sup,\quad \beta>0.
	\label{J_O_inf_sup}
\end{align}
Case $O=0$ corresponds to the minimization or maximization of the entropy itself. Based on this objective, one can define the problem of keeping the thermodynamic observable invariant at the whole time range, steering the entropy to a~given target level, making it follow a~predefined trajectory, etc.
\item For the problem of {\it keeping} the required invariant $S(\rho(t)) \equiv S(\rho_0)$ at the whole time range $[0, T]$, we consider
\begin{align}
	J_1(c) = (S(\rho(T)) - S(\rho_0))^2 + 
	P \int\limits_0^T (S(\rho(t)) - S(\rho_0))^2 dt \to \inf,  
	\label{J1_inf}
\end{align}
where the penalty coefficient~$P>0$ and the final time~$T$ are fixed. Although~one can expect such a~case that making the integral close to zero does not provide $S(\rho(t)) \approx S(\rho_0)$ at the whole $[0,T]$; however, (\ref{J1_inf}) is of interest, because, first, it can be useful and, second, it is appropriate for the described below gradient approach (GPMs). Moreover, as~a~variant, one can formulate the problem
\begin{align} 
	J_2(c) = \max\limits_{\{t_1 > 0, \,\dots,\, t_k, \,\dots,\, t_M = T\}} \left| S(\rho(t_k)) - S(\rho_0) \right| \to \inf,
	\label{J2_inf}
\end{align}
which is considered below together with piecewise linear controls and~GA.
\item For the problem of {\it steering} the von Neumann entropy to a~given target value~$S_{\rm tar}$, we~consider
\begin{align} 
	J_3(c) = (S(\rho(T)) - S_{\rm tar})^2 \to \inf, \qquad S_{\rm tar} \neq S(\rho_0),
	\label{J3_inf}
\end{align}
where $T$ is fixed, as~necessary for the considered GPMs. 
In extension, one can analyze a~series of such steering problems for various values~$T$ and look 
for such an~approximately minimal $T$ for which the required value $S_{\rm tar}$ is reached. 
\item In addition to the steering problem with $J_3$, we consider the 
pointwise state constraint $S(\rho(t)) \leq \overline{S}$ for a~given~$\overline{S} > S(\rho_0)$  at the whole $[0, T]$ by adding to $J_3$ the integral term, taking into account the constraint:
\begin{align} 
	J_4(c) = (S(\rho(T)) - S_{\rm tar})^2 + P \int\limits_0^T (\max\{S(\rho(t)) - \overline{S}, 0 \})^2 dt \to \inf, \quad P>0.
	\label{J4_inf}
\end{align}
Here, the final time~$T$ and the penalty coefficient $P>0$ are fixed. Moreover, as~a~variant, one can consider non-fixed~$T$ and take into account the state constraint as follows:
\begin{align} 
	J_5(c,T) &= \left|S(\rho(T)) - S_{\rm tar}\right| \nonumber \\
	&\quad + P \max\limits_{\{t_1 > 0, \,\dots,\, t_k, \,\dots,\, t_M = T\}} \left(\max\{ S(\rho(t_{{k}})) - \overline{S}, 0 \}\right) \to \inf, \quad P>0,
	\label{J5_inf}
\end{align}
where $T$ is considered free at a~given range $[T_1, T_2]$. As~for~$J_2$, we consider $J_5$ for piecewise linear controls and perform finite-dimensional optimization using~GA.
\end{itemize} 

For the objective functionals $J_1(c),~J_3(c),~J_4(c)$, below~the GPM-1 and GPM-2 are formulated for the class of bounded piecewise continuous controls. For~a~unified description of the GPMs for these three optimal control problems, we use the following notation:
\begin{align}
\Phi(c) &\quad \text{is } \quad J_1(c) \quad \text{or} \quad J_3(c) \quad \text{or} \quad J_4(c), \nonumber \\
F(\rho) &= \begin{cases}
	(S(\rho) - S(\rho_0))^2, & \text{if} \quad J_1 \quad \text{is used}, \\
	(S(\rho) - S_{\rm tar})^2, & \text{if} \quad J_3\text{~~~or~~~}J_4 \text{~~~is used}, 
\end{cases}
\label{function_F_of_rho} \\
g(\rho) &= \begin{cases}
	0, & \text{if} \quad J_3 \quad \text{is used}, \\
	(S(\rho) - S(\rho_0))^2, & \text{if} \quad J_1 \quad \text{is used}, \\
	(\max\{S(\rho) - \overline{S}, 0 \})^2, & \text{if} \quad J_4 \quad \text{is used}. 
\end{cases}
\label{function_g_of_rho}
\end{align} 

The objective functionals $J_2(c),~J_5(c,T)$, as~it is noted above, we consider with piecewise linear controls. Such a~control~$c$ is determined by control parameters corresponding to a~set of nodes at $[0, T]$. For~example, one can consider a~uniform grid 
{$\{t_1 = 0, \dots, t_s, \dots, t_N = T\}$ with the step $\Delta t = T/N$} and the representation 
\[
u(t) = u^{s} + (t - t_{s})(u^{s + 1} - u^{s})/\Delta t, \quad
n_{j}(t) = n_{j}^{s} + (t - t_{s})(n_{j}^{s + 1} - n_j^{s})/\Delta t, \quad j = 1,2 
\]
that allows introducing the vector of parameters, 
\[
{\bf a} = (a_i)_{i=1}^{3N} = (u^1, \dots, u^N, ~ n_1^1, \dots, n_1^N, ~ n_2^1, \dots, n_2^N),  
\]
satisfying the constraints~$|u^{s}| \leq u_{\rm max}$, $n_{j}^{s} \in [0, n_{\max}]$ for $j = 1, 2$ and $s = 1, \dots, N$, and defining such controls~$u,~n_1,~n_2$. Moreover, as~we show below, it can be useful to define a~more sophisticated class 
of controls by defining $c$ as piecewise linear at a~subset of $[0,T]$ and setting constant (zero) for other times; in such a~way, $c$ is defined not only by~${\bf a}$. Anyway, we have deal with finite-dimensional optimization, where $J_2(c),~J_5(c,T)$ are represented by the corresponding objective functions $q_2({\bf a})$ and $q_5({\bf a},T)$ to be minimized. Moreover, for~these objective functions, one can decide to add regularization in controls, e.g.,~for $J_5$, as~follows:
\begin{align}
q_5({\bf a},T; \gamma) = q_5({\bf a},T) + \gamma_u \max\limits_{1 \leq s \leq N}\{|u^{s}|\} + \gamma_n \left(\max\limits_{1 \leq s \leq N}\{n_1^{s}\} + \max\limits_{1 \leq s \leq N}\{n_2^{s}\}\right) \to \inf, 
\label{obj_functions_with_regularization1}
\end{align}
where the coefficients $\gamma_u,~\gamma_n \geq 0$. Moreover, as~a~variant, for~the parameters, which represent incoherent controls, consider the inequality constraints $|n_{j}^{s + 1} - n_j^{s}| \leq \delta_n^{j}$, {$s = 1, \dots, N-1$, $j = 1, 2$}, where the largest allowed jumps $\delta_n^{j} > 0$, $j = 1, 2$ are predefined, and~taking into account these constraints. E.g., for~$J_2(c)$ and $q_2({\bf a})$, consider
\begin{align}
q_2({\bf a}; \gamma) &= q_2({\bf a}) + \gamma_u \max\limits_{1 \leq s \leq N} 
\{|u^{s}|\} \nonumber \\
&\quad + \gamma_n 
\sum\limits_{j = 1}^2 \max\big\{ \max\limits_{1 \leq s \leq N-1} \big\{ \vert n_{j}^{s + 1} - n_{j}^{s} \vert - \delta_n^{j}, 0 \big\} \big\} \to \inf.
\label{obj_functions_with_regularization2}
\end{align}  
This equation is used to take into account possible bounds on variations of the incoherent~control.

For the objectives, for~which GPMs are used below, e.g.,~for $J_3(c)$, one can add the following regularization term (like to~\cite{MorzhinPechenBulletinIrkutsk2023}, p.~14):
\begin{align}
R(c;\gamma) = \int\limits_0^T \Big( \gamma_u u^2(t) + \gamma_n (n_1(t) + n_2(t)) \Big) dt, 
\quad \gamma_u,~\gamma_n \geq 0.
\label{regularization_in_controls_integral}
\end{align}

\section{Markovian Two-Qubit~System}
\label{section4}

As in~\cite{MorzhinPechenQIP2023, MorzhinPechenBulletinIrkutsk2023}, 
consider, as~a~particular case for~(\ref{Eq:ME2}), an~open two-qubit system whose dynamics are determined by a~GKSL-type master equation which contains coherent and incoherent controls and $H^{u, n}_t=H_S + H_{c(t)}$. Here, we deal with the~following:
\begin{itemize}
\item The system state $\rho(t) : \mathcal{H} \to \mathcal{H}$ as a $4 \times 4$ density matrix (positive semi-definite, $\rho(t) \geq 0$, with~unit trace, ${\rm Tr}\rho(t) = 1$) and a~given initial density matrix~$\rho_0$; 
\item Scalar coherent control $u$, vector incoherent control $n=(n_1, n_2)$, and~the corresponding vector control $c=(u,n)$  considered in this work, in~general, as piecewise continuous functions on $[0, T]$; 
\item $H_S$ being the free Hamiltonian defined below; 
\item The controlled Hamiltonian $H_{c(t)} = \varepsilon H_{{\rm eff}, n(t)} + H_{u(t)}$, consisting of the effective Hamiltonian $H_{{\rm eff}, n(t)}$, which 
represents the Lamb shift and depends on $n(t)$, and~of the Hamiltonian $H_{u(t)} = V u(t)$, which describes interaction of the system with $u(t)$ and contains a~Hermitian matrix~$V$ specified below as in~\cite{MorzhinPechenBulletinIrkutsk2023};
\item $\mathcal{D}_t^n$ being the controlled superoperator
of dissipation, where we consider a~special form of a Lindblad superoperator known in the weak coupling limit (see~\cite{PechenRabitzPRA2006}, etc.);
\item The parameter $\varepsilon > 0$ describing the coupling strength between 
the system and the environment;
\item The system of units with the Planck constant $\hbar=1$.  
\end{itemize} 

The following detailed forms of the Hamiltonians are considered:
\begin{align}
H_S &= H_{S,1} + H_{S,2}, \quad 
H_{S,j} = \frac{\omega_j}{2} W_j, \quad 
W_1 := \sigma_z \otimes \mathbb{I}_2, \quad W_2 := \mathbb{I}_2 \otimes \sigma_z, 
\label{free Hamiltonian} \\
H_{{\rm eff}, n(t)} &= \sum\limits_{j=1}^2 H_{{\rm eff}, n_j(t)}, \quad H_{{\rm eff}, n_j(t)} = \Lambda_j W_j n_j(t), \label{effective_Hamiltonian} \\
H_{u(t)} &= V u(t), \quad V = Q_1 \otimes \mathbb{I}_2 + \mathbb{I}_2 \otimes Q_2, \label{Hamiltonian_inter_u} \\
Q_j &= \sum\limits_{\alpha = x,y,z} \lambda_{\alpha}^j \sigma_{\alpha} = \sin\theta_j \cos\varphi_j \sigma_x + \sin\theta_j \sin\varphi_j \sigma_y + \cos\theta_j \sigma_z, \label{matrices_Q_i}
\end{align}
where $j=1,2$. Here $\sigma_x = \begin{pmatrix}
	0 & 1 \\
	1 & 0
\end{pmatrix}$,
$\sigma_y = \begin{pmatrix}
	0 & -i \\
	i & 0
\end{pmatrix}$, and~$\sigma_z = \begin{pmatrix}
	1 & 0 \\
	0 & -1
\end{pmatrix}$   
are the $X$, $Y$, and~$Z$ Pauli matrices. The free Hamiltonian $H_{S,j}$ contains the transition frequency~$\omega_j$ of the $j$th qubit. The~effective Hamiltonian $H_{{\rm eff}, n(t)}$ represents the Lamb shift which describes shifts in transition frequencies of the qubits under the influence of the environment. The~coefficients~$\Lambda_j > 0$, $j=1,2$ together with $n_j(t)$ describe the influence of the environment on the Lamb shift. In~$H_{u(t)}$, the~unit vectors $\lambda^j := (\lambda_x^j, \lambda_y^j, \lambda_z^j) \in \mathbb{R}^3$, $j=1,2$. Physically, the Hamiltonian can describe either a~pair of two-level atoms in electric fields polarized along the directions $\lambda^j := (\lambda_x^j, \lambda_y^j, \lambda_z^j) \in \mathbb{R}^3$, $j=1,2$, or~two particles with spin 1/2 in magnetic fields along the directions $\lambda^j$. In~this model, the~qubits independently interact with the coherent controls of the same intensity but with different directions determined by vectors $\lambda^j$, so that the interaction Hamiltonian $V$ is the sum of two terms. In~\cite{MorzhinPechenQIP2023}, in~addition to this form, the case when coherent control induces interaction between the qubits was also considered. In~contrast to~\cite{MorzhinPechenBulletinIrkutsk2023}, and this work, the~ articles~\cite{MorzhinPechenQIP2023} consider only the case where $Q_1=Q_2=\sigma_x$, i.e.,~in the present terms, $\theta_j = \pi/2$ and $\varphi_j = 0$, $j=1,2$. 

As in \cite{MorzhinPechenQIP2023, MorzhinPechenBulletinIrkutsk2023}, 
consider the following two-qubit superoperator of dissipation:
\begin{align}
\mathcal{D}_t^n(\rho(t)) &= 
\mathcal{D}_{n(t),1}(\rho(t)) + 
\mathcal{D}_{n(t),2}(\rho(t)), 
\label{dissipator}
\\
\mathcal{D}_{n(t),j}(\rho(t)) &=  \Omega_j (n_j(t) + 1) \left( 2 \sigma^-_j \rho \sigma^+_j - 
\sigma_j^+ \sigma_j^- \rho - \rho \sigma_j^ + \sigma_j^- \right) \nonumber \\
&\quad + \Omega_j n_j(t) \left( 2\sigma^+_j \rho \sigma^-_j - 
\sigma_j^- \sigma_j^+ \rho - \rho \sigma_j^- \sigma_j^+ \right),
\qquad j = 1,2. 
\label{dissipator_j}
\end{align}
The coefficients~$\Omega_j>0$, $j=1,2$ are determined by the system--environment microscopic interaction. The matrices $\sigma_j^{\pm}$ are
\begin{equation}
\label{sigma_pm_j}
\sigma_1^{\pm} = \sigma^{\pm} \otimes \mathbb{I}_2, \qquad 
\sigma_2^{\pm} = \mathbb{I}_2 \otimes \sigma^{\pm} \qquad\text{with}\quad 
\sigma^+ = \begin{pmatrix}
	0 & 0 \\ 1 & 0
\end{pmatrix}, \quad
\sigma^- = \begin{pmatrix}
	0 & 1 \\ 0 & 0
\end{pmatrix}.
\end{equation} 

Incoherent control~$n$ has the physical meaning of the density of particles of the system environment and,~therefore, should be non-negative. Moreover, we consider the parallelepipedal constraints:
\begin{eqnarray}
c(t) = (u(t), n_1(t), n_2(t)) \in [-u_{\max}, u_{\max}] \times [0, n_{\max}]^2= Q , \quad  \text{for all} \quad t \in [0, T],
\label{constraint_on_controls}
\end{eqnarray}
where $u_{\max},~n_{\max} > 0$. The~parameters $\varepsilon$, $\omega_1$, $\omega_2$, $\Lambda_1$, $\Lambda_2$, $\theta_1$, $\theta_2$, $\varphi_1$, $\varphi_2$, $\Omega_1$, $\Omega_2$, $u_{\max}$, $n_{\max}$ are considered fixed when we formulate the~optimal control problems, while modifying some of them alters the quantum dynamics, i.e.,~one can vary them for a~deeper~analysis. 

In this article, the~two-qubit system is considered, in~general, with~piecewise continuous controls. The~described below GPMs operate in theory with such controls, and~the performed computer implementations of GPMs use piecewise linear interpolation for controls. For~the non-differentiable objectives, we consider piecewise linear controls that, in~contrast to piecewise constant controls used in the GRAPE-type method in~\cite{PetruhanovPechen2022}, is~another way of parameterization of controls. 

For such a Markovian two-qubit system, the~corresponding evolution equation for real-valued states was obtained in~\cite{MorzhinPechenQIP2023} and has the form
\begin{align}
	\der{x(t)}{t} = \big(A + B_u u(t) + B_{n_1} n_1(t) + B_{n_2} n_2(t) \big) x(t),
	\quad x(0) = x_{\rho_0}, 
	\label{realificated_Markovian_system}
\end{align}
obtained using the parameterization of the system density matrix,
\begin{align}
\rho = \begin{pmatrix}
	\rho_{1,1} & \rho_{1,2} & \rho_{1,3} & \rho_{1,4} \\
	\rho_{1,2}^{\ast} & \rho_{2,2} & \rho_{2,3} & \rho_{2,4} \\
	\rho_{1,3}^{\ast} & \rho_{2,3}^{\ast} & \rho_{3,3} & \rho_{3,4} \\
	\rho_{1,4}^{\ast} & \rho_{2,4}^{\ast} & \rho_{3,4}^{\ast} & \rho_{4,4} 
\end{pmatrix} 
= \begin{pmatrix}
	x_1 & x_2 + i x_3 & x_4 + i x_5 & x_6 + i x_7 \\
	x_2 - i x_3 & x_8 & x_9 + i x_{10} & x_{11} + i x_{12} \\
	x_4 - i x_5 & x_9 - i x_{10} & x_{13} & x_{14} + i x_{15} \\
	x_6 - i x_7 & x_{11} - i x_{12} & x_{14} - i x_{15} & x_{16} 
	\label{parameterization_of_rho}
\end{pmatrix}.
\end{align} 

To analyze the dynamics of each qubit  separately, we consider the reduced density matrices $\rho^j \in  \mathbb{C}^{2\times2}$, $j=1,2$, and~the corresponding Bloch vectors for the two qubits
\begin{align}
	\rho^1 = \mathrm{Tr}_{{\cal H}_2} \rho = \sum_{k = 1}^2 (\mathbb{I}_2 \otimes \langle k |) \rho (\mathbb{I}_2 \otimes | k\rangle ), \qquad 
	\rho^2 = \mathrm{Tr}_{{\cal H}_1} \rho = \sum_{k = 1}^2 (\langle k |  \otimes \mathbb{I}_2) \rho (|k\rangle \otimes  \mathbb{I}_2),
	\label{reduced_density_matrices_for_qubits_vs_t}
\end{align} 
where $| k\rangle$ are basis vectors in $\mathcal{H}_1$ and $\mathcal{H}_2$. Because~the density matrix of a~qubit can be bijectively mapped to the Bloch ball (in $\mathbb{R}^3$, this ball is centered in the point~$(0,0,0)$ and has the unit radius), consider Bloch vectors $r^j = (r_x^j, r_y^j, r_z^j)$ where 
$r_{\alpha}^j = \mathrm{Tr}(\rho^j \sigma_{\alpha})$, $\alpha \in \{x, y, z\}$, 
$|r^j|\le 1$, $j=1,2$. In~terms of parameterization~(\ref{parameterization_of_rho}), one has:
\begin{align}
	r^1 &= \left( 2 (x_4 + x_{11}), \quad
	-2 (x_5 + x_{12}), \quad 
	x_1 + x_8 - x_{13} - x_{16} \right), 
	\label{qubit1_Bloch_vector}
	\\
	r^2 &= \left( 2 (x_2 + x_{14}), \quad 
	-2 (x_3 + x_{15}), \quad
	x_1 - x_8 + x_{13} - x_{16} \right).
	\label{qubit2_Bloch_vector}
\end{align}
Reduced density matrices are  
$\rho^j = \frac{1}{2} \begin{pmatrix}
	1 + r_z^j & r_x^j - i \, r_y^j \\
	r_x^j + i \, r_y^j & 1 - r_z^j
\end{pmatrix}$, 
$j = 1,2$. Further, for~density matrices 
$\rho^1(t)$ and $\rho^2(t)$ vs $t \in [0,T]$, we consider their von~Neumann entropies, 
i.e., $S(\rho^j(t)) = - \mathrm{Tr} \left(\rho^j(t) \log \rho^j(t) \right)$,
$j=1,2$, and~the sum $S(\rho^1(t)) + S(\rho^2(t))$. The behavior of these quantities in the numerical experiments is shown below in Figures~\ref{fig1}b and \ref{fig2}c,f,i.

\begin{figure}[h!]
\centering 
\includegraphics[width=\linewidth]{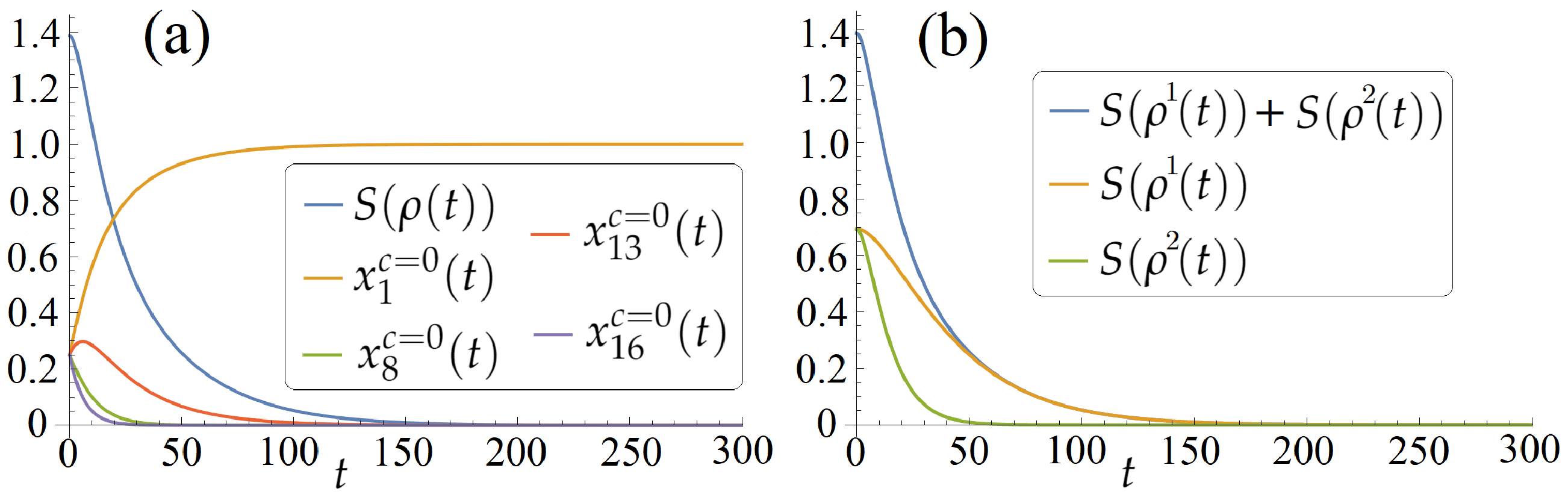}  
\caption{For the initial state $\rho_0 = \frac{1}{4}\mathbb{I}_4$ and the control $c=0$: (\textbf{a})~the von Neumann entropy $S(\rho(t))$ and $x_j^{c=0}(t)$, $j=1,8,13,16$, i.e.,~the diagonal elements of the diagonal $\rho(t)$, vs $t \in [0, T=300]$ (in this case, the~entropy steers from the largest value $\log 4 \approx 1.39$ to zero, indicating the system's state purification and minimization of~$S(\rho(T)$); (\textbf{b})~the entropies $S(\rho^1(t))$ and $S(\rho^2(t))$ for the first and second qubits, correspondingly, and~the sum $S(\rho^1(t))+S(\rho^2(t))$, vs $t \in [0, T=300]$, steer to zero}.\label{fig1}
\end{figure} 
\vspace{-12pt}

\begin{figure}[h!]
\centering 
\includegraphics[width=\linewidth]{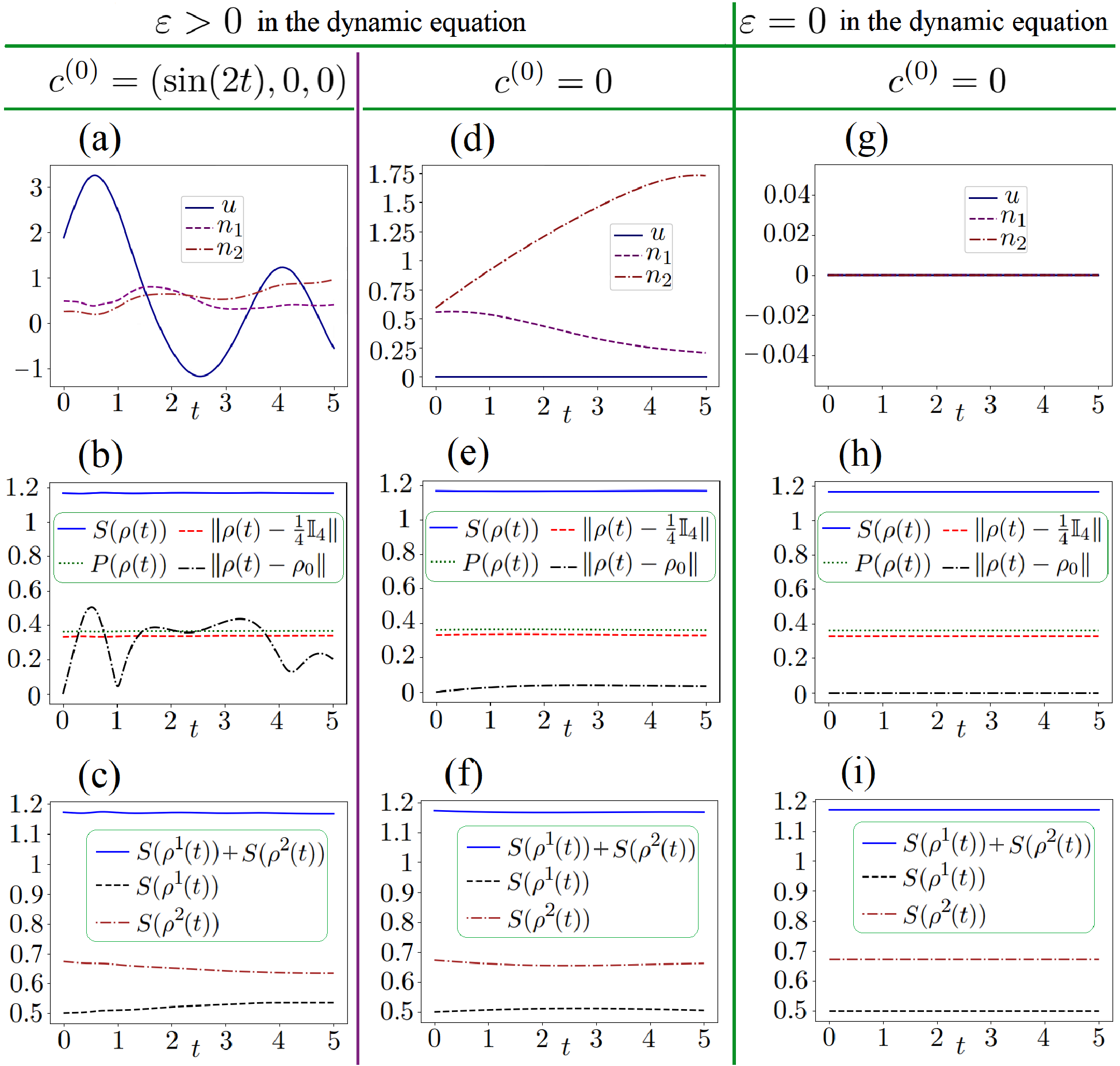}  
\caption{For the problem of keeping the invariant $S(\rho(t)) \equiv S(\rho_0)$ at the whole $[0,T=5]$. Problem~(\ref{J1_inf}) and GPM-2 are used: (1)~the subfigures (\textbf{a}--\textbf{c}) shows the results for $\varepsilon = 0.1$ and $c^{(0)}=(\sin(2t),0,0)$; (2)~the subfigures (\textbf{d}--\textbf{f}) shows the results for $\varepsilon = 0.1$ and $c^{(0)}=0$; (3)~the subfigures (\textbf{g}--\textbf{i}) shows the results for $\varepsilon = 0$ (i.e., without taking into account the Lamb shift and the dissipator) and $c^{(0)}=0$. The subfigures (\textbf{a,d,g}) show the obtained controls; for these controls, the subfigures (\textbf{b,e,h}) and (\textbf{c,f,i}) show, correspondingly, the two-qubit system characteristics ($S(\rho(t))$, etc.) vs~$t$ and the entropies $S(\rho^1(t))$, $S(\rho^2(t))$, their sums vs $t$. In~the cases related to the subfigures~(\textbf{c},\textbf{f}), we see that for each qubit its entropy is not constant. \label{fig2}} 
\end{figure} 

\section{Numerical Optimization Tools: Markovian Two-Qubit~Case}
\label{section5}

\subsection{Gradient-Based Optimization Approach for the Problems with $J_1,~J_3,~J_4$}

\subsubsection{Pontryagin Function and Krotov~Lagrangian}

According to the theory of optimal control (e.g., \cite{PontryaginBook_1962}), for~the unified optimal control problem with $\Phi(c)$ representing $J_1,~J_3,~J_4$, the~Pontryagin function is   
\begin{align*}
h(\chi,\rho,c) &= \left\langle \chi, -i[H_c, \rho] + \varepsilon  \mathcal{D}_n(\rho) \right\rangle  - P\,g(\rho) - \gamma_u u^2 - \gamma_n (n_1 + n_2) \nonumber \\
&\quad = \langle \mathcal{K}^c (\chi,\rho), c \rangle  - \gamma_u u^2 - \gamma_n (n_1 + n_2) + \overline{h}(\chi,\rho),
\end{align*} 
where $\chi$ and $\rho$ are $4 \times 4$ density matrices; $c = (u, n_1, n_2) \in \mathbb{R}^3$; the functions
\begin{align}
&\mathcal{K}^c = (\mathcal{K}^u, \mathcal{K}^{n_1}, \mathcal{K}^{n_2}), \quad \mathcal{K}^u(\chi,\rho) = \left\langle \chi, -i[V, \rho] \right\rangle , 
\label{switching_functions_Kc_Ku} \\
&\mathcal{K}^{n_j}(\chi,\rho) = \Big\langle \chi, -i[\Lambda_j W_j, \rho] + 
\varepsilon \Omega_j\Big(2 \sigma_j^- \rho \sigma_j^+ + 2\sigma_j^+ \rho \sigma_j^- - \big\{ \mathbb{I}_4, \rho \big\} \Big) \Big\rangle , \quad j = 1,2 \label{switching_functions_Kn_j}
\end{align}
(the $4\times 4$ identity matrix $\mathbb{I}_4$ appears in $\mathcal{K}^{n_j}$ since $\sigma_j^+ \sigma_j^- + \sigma_j^- \sigma_j^+ = \mathbb{I}_4$), $j=1,2$; the term
\[
\overline{h}(\chi,\rho) = \Big\langle \chi, -i [H_S, \rho] + \varepsilon \sum\limits_{j=1}^2 \Omega_j \big(2 \sigma_j^- \rho \sigma_j^+ - \big\{ \sigma_j^+ \sigma_j^-, \rho \big\} \big) \Big\rangle  - P\,g(\rho).
\]

As the Introduction notes, various Krotov-type iterative methods are used in quantum optimal control. In~this article, we do not use any Krotov-type method, but we use the Krotov Lagrangian, which is the following for the unified problem:
\begin{align*}
L(c, \rho) &= G(\rho(T)) - \int_0^T R(t, \rho(t), c(t))dt, \\
G(\rho(T)) &= F(\rho(T)) + \langle \chi(T), \rho(T) \rangle  - \langle \chi(0), \rho_0 \rangle , \\
R(t,\rho,c) &= \left\langle \chi(t), -i[H_c, \rho] + \varepsilon \mathcal{L}^D_n(\rho) \right\rangle  + \langle \dot \chi(t), \rho\rangle  - P\,g(\rho) - \gamma_u u^2 - \gamma_n (n_1 + n_2).
\end{align*}
The function $\chi$ is defined in the next subsection as the solution of the adjoint system also defined below. For~each admissible control~$c$, the~values of the Krotov Lagrangian and $\Phi(c)$ coincide, as~in the general V.F.~Krotov theory~\cite{KrotovBook1996}.

\subsubsection{Unified Adjoint System and~Gradient}

Consider the increment of $L$ at admissible controls $c, c^{(k)}$ (for the further consideration, we introduce $k \geq 0$ as an~iteration index): 
\begin{align}
\label{incrementKrotovLagrangian}
L(c, \rho) - L(c^{(k)}, \rho^{(k)}) &= G(\rho(T)) - G(\rho^{(k)}(T)) \nonumber \\
&\quad - \int_0^T (R(t, \rho(t), c(t)) -
R(t, \rho^{(k)}(t), c^{(k)}(t)))dt,
\end{align}
where the control process $(c^{(k)}, \rho^{(k)})$ is~known. 

By analogy with \cite{KrotovBook1996} (pp.~239--240) in the theory of optimal control, here for the increment~(\ref{incrementKrotovLagrangian}), we consider the first-order Taylor expansions for $G,~R$. At~admissible controls $c,~c^{(k)}$, this gives the representation 
\begin{align*} 
\Phi(c) - \Phi(c^{(k)}) &= 
\Big\langle \der{}{\rho} G(\rho^{(k)}(T), \rho(T) - \rho^{(k)}(T) \Big\rangle  \nonumber \\
&\quad - \int_0^T \big\langle \frac{\partial}{\partial \rho} R(t, \rho^{(k)}(t), c^{(k)}(t)), \rho(t) - \rho^{(k)}(t) \big\rangle dt \nonumber \\
&\quad - \int_0^T \big\langle \frac{\partial}{\partial c} R(t, \rho^{(k)}(t), c^{(k)}(t)), c(t) - c^{(k)}(t) \big\rangle_{E^3} dt + r.
\end{align*}
Here, the notations with the derivatives mean that we initially find these derivatives with respect to $\rho$ or $(\rho,c)$, and after that, we substitute $\rho = \rho^{(k)}(T)$, etc.; $r$~is the corresponding residual. Setting the derivatives $\der{}{\rho} G(\rho^{(k)}(T)$ and $\dfrac{\partial}{\partial \rho} R(t, \rho^{(k)}(t), c^{(k)}(t))$ to be zero gives the {\it adjoint system} which defines the function $\chi^{(k)}$ as detailed below. As~the result, the~increment formula for the unified objective $\Phi(c)$ has the form 
\begin{align}
&\Phi(c) - \Phi(c^{(k)}) =  - \int_0^T \Big\langle \frac{\partial}{\partial c} R(t, \rho^{(k)}(t), \, c^{(k)}(t)), c(t) - c^{(k)}(t) \Big\rangle_{E^3} dt + r, 
\label{increment_formula_1} \\
&\frac{\partial}{\partial c} R(t, \rho^{(k)}(t), c^{(k)}(t)) = 
\frac{\partial}{\partial c} h(\chi^{(k)}(t), \rho^{(k)}(t), c^{(k)}(t)) = \mathcal{K}^c(\chi^{(k)}(t), \rho^{(k)}(t))  \nonumber \\
&\quad - \gamma_u (u^{(k)}(t))^2 - \gamma_n (n^{(k)}_1(t) + n^{(k)}_2(t)). \nonumber
\end{align} 

The differentiation of the unified function~$F$ is needed to obtain the condition for the final co-state $\chi^{(k)}(T)$, i.e.,~the transversality condition; for differentiation of~$F$, it is needed to consider the various forms of~$F$ shown in~(\ref{function_F_of_rho}). For~differentiation of~$R$, it is needed to consider the various forms of~$g(\rho)$ shown in~(\ref{function_g_of_rho}). Using the matrix differential calculus (e.g.,~\cite{MatrixCookbook2012}), for~the  problems the following derivatives are found:
\begin{align*}
\der{S(\rho)}{\rho} &= -\log \rho - \mathbb{I}_{\dim \mathcal{H}=4}, \nonumber \\
\der{F(\rho)}{\rho} &= \begin{cases}
	\der{}{\rho} (S(\rho) - S(\rho_0))^2, & \text{if} \quad J_1 \quad \text{is used}, \\
	\der{}{\rho} (S(\rho) - \overline{S})^2, & \text{if} \quad J_3\text{~~~or~~~}J_4 \text{~~~is used}, 
\end{cases} \nonumber \\
&= -2(\log\rho + \mathbb{I}_4) \begin{cases}
	S(\rho) - S(\rho_0), & \text{if} \quad J_1 \quad \text{is used}, \\
	S(\rho) - \overline{S}, & \text{if} \quad J_3\text{~~~or~~~}J_4 \text{~~~is used}, 
\end{cases}  \\
\der{g(\rho)}{\rho} &= 
\begin{cases}
	0, & \text{if} \quad J_3 \quad \text{is used}, \\
	\der{}{\rho} (S(\rho) - S(\rho_0))^2, & \text{if} \quad J_1 \quad \text{is used}, \\
	\der{}{\rho} (\max\{S(\rho - \overline{S}, 0 \})^2, & \text{if} \quad J_4 \quad \text{is used} \end{cases} \nonumber \\
&= 
-2 (\log\rho + \mathbb{I}_4) 
\begin{cases}
	0, & \text{if} \quad J_3 \quad \text{is used}, \\
	S(\rho) - S(\rho_0), & \text{if} \quad J_1 \quad \text{is used}, \\
	\max\{ S(\rho) - \overline{S}, 0 \}, & \text{if} \quad J_4 \quad \text{is used.} 
\end{cases}  
\end{align*} 

To compute the derivative $\dfrac{\partial}{\partial \rho} R(t, \rho^{(k)}(t), c^{(k)}(t))$, one needs to operate with the right-hand side of the system~(\ref{Eq:ME2}) and take into account the corresponding properties such that the anti-commutativity property of commutator and cyclic permutation of matrices under trace. In~this regard, and using the given formulas above for $\der{g(\rho)}{\rho}$, we, as~a result, obtain the adjoint system shown below in Proposition~\ref{proposition_1}. This adjoint system contains the following superoperator acting on $\chi^{(k)}(t)$ (this superoperator is the same as derived in~\cite{MorzhinPechenBulletinIrkutsk2023}):  
\begin{align*}
\mathcal{D}^{\dagger}_{n^{(k)}(t)}(\chi^{(k)}(t)) = \sum\limits_{j=1}^2 \Big[ \Omega_j \left( n^{(k)}_j(t) + 1 \right) \Big(2 \sigma_j^+ \chi^{(k)}(t) \sigma_j^- -  \left\{ \sigma_j^+ \sigma_j^-, \chi^{(k)}(t) \right\}\Big) \nonumber \\    
+ \Omega_j n^{(k)}_j(t) \Big(2 \sigma_j^- \chi^{(k)}(t) \sigma_j^+  - \left\{ \sigma_j^- \sigma_j^+, \chi^{(k)}(t) \right\} \Big) \Big],
\end{align*}
where ``$\dagger$'' reflects that 
$\langle \chi^{(k)}(t), T_1(\rho(t)) - T_1(\rho^{(k)}(t)) \rangle  = \langle T_1^{\dagger}(\chi^{(k)}(t)), \rho(t) - \rho^{(k)}(t) \rangle$ and also $\langle \chi^{(k)}(t), T_2(\rho(t)) - T_2(\rho^{(k)}(t)) \rangle  = \langle T_2^{\dagger}(\chi^{(k)}(t)), \rho(t) - \rho^{(k)}(t)\rangle$ with the operators $T_1 := 2\sigma_j^- \boldsymbol{\cdot} \sigma_j^+$, $T_2 := 2\sigma_j^+ \boldsymbol{\cdot} \sigma_j^-$ and $(\sigma_j^+)^{\top}=\sigma_j^-$, $(\sigma_j^-)^{\top}=\sigma_j^+$; note that  $\sigma_j^+ \sigma_j^-$, $\sigma_j^- \sigma_j^+$ are~Hermitian. 

\begin{proposition} (Adjoint system).  
\label{proposition_1}
For the Markovian two-qubit case of the system (\ref{Eq:ME2}) and the unified objective functional $\Phi(c)$ containing the unified terminant $F(\rho(T))$ and integrand $g(\rho(t))$, the~adjoint system has the following form:
\begin{align}
	\der{\chi^{(k)}(t)}{t} &= -i[H_{c^{(k)}(t)}, \chi^{(k)}(t)] - \varepsilon \mathcal{D}^{\dagger}_{n^{(k)}(t)}(\chi^{(k)}(t)) - P\, \der{g(\rho^{(k)}(t))}{\rho},
	\label{adjoint_system_dyn_eq} \\
	\chi^{(k)}(T) &= -\der{F(\rho^{(k)}(T))}{\rho}. 
	\label{transversality_condition}
\end{align}
\end{proposition}

If the adjoint system is used with taking into account one of the two pointwise state constraints, then the system depends on~$\rho^{(k)}$. Anyway, the~adjoint system is linear in co-state~$\chi^{(k)}$. This system is solved backward in time. In~view of~(\ref{increment_formula_1}) and like the formula~(3.16) in~\cite{MorzhinPechenBulletinIrkutsk2023}, we consider the gradient of the unified~objective.

\begin{proposition} (Gradient). 
\label{proposition_2}
For the Markovian two-qubit case of the system (\ref{Eq:ME2}) and the unified objective functional $\Phi(c)$, the~corresponding gradient at a~given admissible control $c^{(k)}$ has the form
\begin{align}
	{\rm grad}\, \Phi(c^{(k)})(t) &= \Big(-\mathcal{K}^u(\chi^{(k)}(t), \rho^{(k)}(t)) + 2 \gamma_u u^{(k)}(t), \, \nonumber \\
	&\quad\quad -\mathcal{K}^{n_j}(\chi^{(k)}(t), \rho^{(k)}(t))
	+ \gamma_n, \quad j = 1,2 \Big), \qquad t \in [0, T].
	\label{gradient_of_unified_objective_Phi}
\end{align}
Here $\rho^{(k)}$ is the solution of the {Markovian case of the} system~(\ref{Eq:ME2}) with control~$c^{(k)}$, while  $\chi^{(k)}$ is the solution of the adjoint system~(\ref{adjoint_system_dyn_eq}), (\ref{transversality_condition}) with the control process $(\rho^{(k)}, c^{(k)})$; the vector function $\mathcal{K}^c(\chi,\rho)$ defined in~(\ref{switching_functions_Kc_Ku}),~(\ref{switching_functions_Kn_j}) is used with these solutions. 
\end{proposition}

In general, the~formula~(\ref{gradient_of_unified_objective_Phi}) for the unified gradient reminds us, e.g.,~of the gradient formula~(2.5.29) given in Reference~\cite{PolakBook1971} on the theory of optimal control with real-valued~states. 

\subsubsection{Projection Form of the~PMP}

Following the projection form of the PMP known in the theory of optimal control~(e.g., see~\cite{Srochko_Mamonova_2001}) and also its use in quantum control~\cite{BuldaevMathematics2022}, below~such a projection form of the PMP is~formulated.

\begin{proposition} (Projection form of the differential version of the PMP for the unified problem with the objective~$\Phi(c)$). 
\label{proposition_3}
For the Markovian two-qubit case of the system (\ref{Eq:ME2}) and the unified objective functional $\Phi(c)$ with piecewise continuous controls satisfying~(\ref{constraint_on_controls}) for a~fixed final time~$T>0$, if~an~admissible control $\widehat{c} = (\widehat{u}, \widehat{n}_1, \widehat{n}_2)$ is a~local minimum point of $\Phi(c)$ to be minimized, then for $\widehat{c}$ there exist such the solutions~$\widehat{\rho}$ and $\widehat{\chi}$ that the pointwise condition
\begin{align}
	\widehat{c}(t) = {\rm Pr}_Q\big(\widehat{c}(t) - \alpha \, {\rm grad}\, \Phi(\widehat{c})(t) \big), \quad t \in [0,T], \quad \alpha > 0, 
	\label{projection_condition_PMP}
\end{align} 
holds and, in~detail, has the form
\begin{align*}
	\widehat{u}(t) &= \begin{cases}
		-u_{\max}, & \widehat{u}(t;\alpha) < -u_{\max}, \\
		u_{\max}, & \widehat{u}(t;\alpha) > u_{\max}, \\
		\widehat{u}(t;\alpha), & |\widehat{u}(t;\alpha)| \leq u_{\max},
	\end{cases} \\
	&\qquad \text{where} \quad \widehat{u}(t;\alpha) =  \widehat{u}(t) + \alpha (\mathcal{K}^u(\widehat{\chi}(t), \widehat{\rho}(t)) - 2\gamma_u \widehat{u}(t)),\\
	\widehat{n}_j(t) &= \begin{cases}
		0, & \widehat{n}_j(t;\alpha) < 0, \\
		n_{\max}, & \widehat{n}_j(t;\alpha) > n_{\max}, \\
		\widehat{n}_j(t;\alpha), & \widehat{n}_j(t;\alpha) \in [0, n_{\max}],
	\end{cases}\\
	&\qquad \text{where} \quad \widehat{n}_j(t;\alpha) = 
	\widehat{n}_j(t) + \alpha (\mathcal{K}^{n_j}(\widehat{\chi}(t), \widehat{\rho}(t)) - \gamma_n), \quad j = 1,2.  
\end{align*} 
\end{proposition}

\subsubsection{One- and Two-Step Gradient Projection~Methods}

In the theory of optimal control, there are various forms of GPM-1 operating with control functions~(e.g., see in~\cite{LevitinPolyak1966, DemyanovRubinovBook1970, FedorenkoBook1978}). In~quantum control, for~example, work~\cite{Morzhin_Pechen_IJTP_2021} exploits GPM-1, which uses two algorithmic parameters (coefficient~$\alpha$ for the gradient of the considered in that article objective functional and parameter~$\theta \in [0,1]$ of the convex combination between the given control~$c^{(k)}$ and  depending on the $\alpha$ projection form for constructing $c^{(k+1)}$) and a~scheme of one-dimensional optimization with respect to~$\theta$ at each iteration, to search for the best variation of $c^{(k)}$ in the sense of the best decreasing objective. In~contrast to~\cite{Morzhin_Pechen_IJTP_2021}, this article considers GPM-1 without the aforementioned convex combination and with a~fixed~$\alpha$ at the whole set of iterations. The~considered GPM-2 is based on the heavy-ball method (see the works~\cite{PolyakUSSRComputMathMathPhys1964, PolyakBook1987}), its projection version~\cite{AntipinDifferEqu1994, Vasilev_Amochkina_Nedich_1996} and the recent papers~\cite{MorzhinPechenQIP2023, MorzhinPechenBulletinIrkutsk2023}, where the corresponding GPM-2 adaptations are used for quantum~control. 

For the unified optimal control problem and a~given admissible initial guess $c^{(0)}$, consider the following GPMs iterative processes operating in the functional space of~controls.
\begin{itemize}
\item GPM-1. The iteration process in the vector form is as follows and is reminiscent of~(\ref{projection_condition_PMP}):
\begin{align}
	c^{(k+1)}(t) = {\rm Pr}_Q\big(c^{(k)}(t) - \alpha\, {\rm grad}\, \Phi(c^{(k)})(t) \big), \quad \alpha > 0, \quad k \geq 0.
	\label{iteration_formula_GPM1}
\end{align} 
In detail, we have
\begin{align*}
	u^{(k+1)}(t) &= \begin{cases}
		-u_{\max}, & u^{(k)}(t;\alpha) < -u_{\max}, \\
		u_{\max}, & u^{(k)}(t;\alpha) > u_{\max}, \\
		u^{(k)}(t;\alpha), & |u^{(k)}(t;\alpha)| \leq u_{\max},
	\end{cases} \\
	&\qquad \text{where} \quad u^{(k)}(t;\alpha) =  u^{(k)}(t) + \alpha (\mathcal{K}^u(\chi^{(k)}(t), \rho^{(k)}(t)) - 2\gamma_u u^{(k)}(t)),\\
	n_j^{(k+1)}(t) &= \begin{cases}
		0, & n_j^{(k)}(t;\alpha) < 0, \\
		n_{\max}, & n_j^{(k)}(t;\alpha) > n_{\max}, \\
		n_j^{(k)}(t;\alpha), & n_j^{(k)}(t;\alpha) \in [0, n_{\max}],
	\end{cases}\\
	&\qquad \text{where} \quad n_j^{(k)}(t;\alpha) = 
	n_j^{(k)}(t) + \alpha (\mathcal{K}^{n_j}(\chi^{(k)}(t), \rho^{(k)}(t)) - \gamma_n), \quad j = 1,2;  
\end{align*} 

\item GPM-2. The iteration process in the vector form is as follows:
\begin{align} 
	c^{(k+1)}(t) &= {\rm Pr}_Q\big(c^{(k)}(t) - \alpha\, {\rm grad}\, \Phi(c^{(k)})(t)  \nonumber \\
	&\quad + \beta (c^{(k)}(t) - c^{(k-1)}(t)) \big), \quad \alpha, \beta > 0, \quad k \geq 1, 
	\label{iteration_formula_GPM2} 
\end{align}
where $c^{(1)}$ is obtained using GPM-1 for a~given initial guess $c^{(0)}$.
\end{itemize}
Here, the algorithmic parameters $\alpha, \beta >0$ are fixed for all iterations. One may consider this, on~the one hand, as~a~drawback, because~we do not try to effectively variate these parameters, and, on the other hand, as~a~simpler case for the analysis. Moreover, here, relying on the various known computational facts about the heavy-ball method (e.g., see~\cite{SutskeverMartensDahlHinton2013, TensorFlow_MomentumOptimizer}), we take $\beta \in (0, 1)$ and more likely $\beta = 0.8, 0.9$ in  GPM-2, but~not $\beta = 10$, etc. {\tt TensorFlow MomentumOptimizer} \cite{TensorFlow_MomentumOptimizer} under the setting {\tt use\_nesterov = False} represents the heavy-ball method, where the parameter is 0.9 by~default. 

\subsection{Zeroth-Order Stochastic Optimization for the Problems with~$J_2,~J_5$}

GA belongs to zeroth-order stochastic tools, such as differential evolution, simulated annealing, particle-swarm optimization, sparrow search algorithm, etc., whose stochastic behavior models try to find a~global minimizer of an~objective function without its gradient due to these behavior models. In~this article, the~GA implementation~\cite{Solgi_Genetic_Algorithm_Python} has been adjusted for the problems with the objectives~$J_2,~J_5$.  

When a GA realization works with large $u_{\max},~n_{\max}$, then one can expect that the algorithm may miss a~closer-to-optimal point, which is in a~smaller subdomain. Because~of the stochastic nature of GA, one can expect that, for~the same optimization problem, the~results of different trials of the GA may differ significantly even with the same deterministic settings (mutation probability, etc.). That is why one can perform---for the same optimization problem---several trials of the GA and then select the lowest computed value of the objective over the trials. However, e.g.,~if we consider the keeping problem~(\ref{J2_inf}) with regularization in controls and consider $J_2$ as sufficiently close to zero, and the profiles in the computed controls are acceptable, then it is not needed to perform more trials of the GA, because~we know that zero is the lower bound for~$J_2$. 

\section{Analytical and Numerical Analysis: Markovian Two-Qubit~Case}
\label{section6}

In the numerical experiments, the~following values of the system parameters are used:
\begin{align}  
&\omega_1 = 1, \quad \omega_2 = 0.5, \quad 
\Lambda_1 = 0.3, \quad \Lambda_2 = 0.5, \quad \Omega_1 = 0.2, \quad \Omega_2 = 0.6, \quad \varepsilon = 0.1, \nonumber \\
& \varphi_1 = \pi/4, \quad \varphi_2 = \pi/3, \quad \theta_1 = \pi/3, \quad \theta_2 = \pi/4.  
\label{values_of_system_parameters}
\end{align} 
(except for Case~3 in Section~\ref{subsection_6_2}, where for comparison, we set $\varepsilon = 0$). All the parameters are expressed in the relative units of free oscillation of the first qubit, which has period $T_1=2\pi$. Free oscillations of the second qubit have period $T_2=2T_1$. The decoherence rate is by the order of magnitude smaller than the oscillations of the first qubit. The difference between the qubit's free transition frequencies may occur twice, for example, in superconducting qubits. The system-environment coupling is determined by the parameter $\varepsilon$. This parameter specifies the (uncontrolled) decoherence rate, i.e., the rate of decoherence when $u=0$ and $n\equiv 0$).  Generally, the decoherence rate is several orders of magnitude smaller than the rate of free dynamics. In this study, we focus on cases where the decoherence rate is an order of magnitude slower than the free dynamics. 

In the computer realizations (in Python) of GPM-1 and GPM-2, piecewise linear interpolation of controls $u,~n_1,~n_2$ is used at a~uniform grid introduced over~$[0,T]$ with~$M$ subintervals, i.e.,~with $M+1$ time instances. To~solve the considered ODEs, {\tt solve\_ivp} from {\tt SciPy} is~used. 

\subsection{Results on the von Neumann Entropy under Zero Coherent and Incoherent~Controls}
\label{subsection_6.1}

If one takes $c = 0$, then (\ref{realificated_Markovian_system}) becomes ${\der{x^{c=0}}{t}} = A x^{c=0}$, $x^{c=0}(0) = x_{\rho_0}$ whose solution is $x^{c=0}(t) = e^{At} x_{\rho_0}$. For~the parameterized initial density matrix $\rho_0 = {\rm diag}(a_1, a_2, a_3, a_4)$ (s.t. $a_j \geq 0$, $j = 1,2,3,4$, $\sum\limits_{j=1}^4 a_j = 1$) and the corresponding initial state $x_{\rho_0} = (a_1,~\text{six zeros},~a_2,~\text{four zeros},~a_3,~0,~0,~a_4)$, as~Reference~\cite{MorzhinPechenQIP2023} shows,~system~(\ref{realificated_Markovian_system}) for $c = 0$ has the following exact solution: 
\begin{align}
x_1^{c=0}(t) &=   a_1 + a_2 - a_2 e^{-2 \varepsilon\Omega_2 t} + e^{-2 \varepsilon(\Omega_1 + \Omega_2) t} 
(e^{ 2 \varepsilon\Omega_1 t} - 1) (a_3 e^{2\varepsilon \Omega_2  t} + 
a_4 (e^{2 \varepsilon \Omega_2 t} - 1)), \nonumber \\
x_8^{c=0}(t) &=   e^{-2  \varepsilon\Omega_2 t} (a_2 + a_4 - a_4 e^{-2 \varepsilon\Omega_1 t}), \quad
x_{13}^{c=0}(t) = e^{-2  \varepsilon \Omega_1 t} (a_3 + a_4 - a_4 e^{-2  \varepsilon \Omega_2 t}), \nonumber \\ 
x_{16}^{c=0}(t) &=   a_4 e^{-2 \varepsilon(\Omega_1 + \Omega_2) t}, \quad
x_j^{c=0}(t) = 0, \quad j \in \overline{1,16} \setminus \{1, 8, 13, 16\}, \quad {t \geq 0}. \qquad
\label{analytical_solution_quantum_system_x_under_zero_controls}
\end{align}
The corresponding density matrix $\rho$ is diagonal. Then the final von~Neumann entropy is
\begin{align}
S(\rho(T)) = -\sum_{{x_j^{c=0}(T) \neq 0},~j=1,8,13,16} x_j^{c=0}(T) \log x_j^{c=0}(T).
\label{S_rho_T_particular_case_zero_controls}
\end{align}  
Using~(\ref{qubit1_Bloch_vector}),~(\ref{qubit2_Bloch_vector}), we obtain for the Bloch vectors: 
\begin{align*}
	r^1(t) = \left(r_x^1(t), ~ r_y^1(t), ~ r_z^1(t) \right) = \left(0, ~ 0, ~ x_1^{c=0}(t) + x_8^{c=0}(t) - x_{13}^{c=0}(t) - x_{16}^{c=0}(t) \right), \\
	r^2(t) = \left(r_x^2(t), ~ r_y^2(t), ~ r_z^2(t) \right) = \left(0, ~ 0, ~ x_1^{c=0}(t) - x_8^{c=0}(t) + x_{13}^{c=0}(t) - x_{16}^{c=0}(t) \right). 
\end{align*}
Thus, the~$j$th reduced density matrix is also diagonal, $\rho^j(t) = \frac{1}{2} 
\begin{pmatrix}
	1 + r_z^j(t) & 0 \\
	0 & 1 - r_z^j(t)
\end{pmatrix}$,  
and we have
$S(\rho^j(t)) = \begin{cases}
	-\frac{1+r_z^j(t)}{2}\log\frac{1+r_z^j(t)}{2} - 
	\frac{1-r_z^j(t)}{2}\log\frac{1-r_z^j(t)}{2}, & \text{if}~~ r^j_z(t) \not \in \{\pm 1\}, \\
	0, & \text{if otherwise}. 
\end{cases}$

{\bf Case 1:} $\rho_0 = \frac{1}{4}\mathbb{I}_4$ ($a_1 = a_2 = a_3 = a_4 = \frac{1}{4}$), i.e.,~the completely mixed quantum state whose von~Neumann entropy is the largest among $4 \times 4$ density matrices. Using~(\ref{S_rho_T_particular_case_zero_controls}), for~(\ref{values_of_system_parameters}) and $T = 50, \, 200, \, 250$, we obtain, correspondingly, $S(\rho(T)) \approx 0.2571, \, 0.0016, \, 0.0003$. For~a~sufficiently large~$T$, this steering allows the purification of the system states with~good quality. This corresponds to the problem of minimizing the objective functional $J_0(c) = S(\rho(T)) \to \inf$ that relates to~(\ref{J_O_inf_sup}). We see that in the considered case, the~purification goal is achieved using the system-free evolution, i.e.,~without any non-trivial control~$c$. Figure~\ref{fig1} shows $x_j^{c=0}(t)$, $j = 1, 8, 13, 16$, and~$S(\rho(t))$ computed via~(\ref{S_rho_T_particular_case_zero_controls}) vs~$t \in [0,T=300]$. We see that approximately $x_1^{c=0}$ steers to~1, while $x_8^{c=0}$, $x_{13}^{c=0}$, and~$x_{16}^{c=0}$ steer to zero. This means that the system approximately steers to the pure state~$\rho = {\rm diag}(1,0,0,0)$.

{\bf Case 2:} $\rho_0 = {\rm diag}\big(\frac{1}{2}, \frac{3}{10}, \frac{1}{10}, \frac{1}{10} \big)$, i.e.,~a~mixed quantum state. If~we take Formula~ (\ref{analytical_solution_quantum_system_x_under_zero_controls}) with $\varepsilon = 0$, then we have $x_1^{c=0}(t) \equiv \frac{1}{2}$, $x_8^{c=0}(t) \equiv \frac{3}{10}$, $x_{13}^{c=0}(t) \equiv \frac{1}{10}$, and~$x_{16}^{c=0}(t) \equiv \frac{1}{10}$ for any~$t \geq 0$. For~any time, this particular dynamic system does not leave the state $\rho_0$ ($x_{\rho_0}$)---this is a~{\it singular point} of the system vector field. This analytical finding relates with one of the considered below cases for the keeping problem (and with the right column of the subfigures in Figure~\ref{fig2}) analyzed in the next~subsection. 

\subsection{The Problem of Keeping the Initial Entropy $S(\rho_0)$} 
\label{subsection_6_2}

Consider the initial state  
$\rho_0 = {\rm diag}\big(\frac{1}{2}, \frac{3}{10}, \frac{1}{10}, \frac{1}{10} \big)$ with $S(\rho_0) \approx 1.168$ and the problem of keeping the von Neumann entropy~$S(\rho(t))$ at the level~$S(\rho_0)$ at the whole $[0,T=5]$. 

\subsubsection{Using the Problem~(\ref{J1_inf}) and GPM} 

Set the coefficient~$P = 0.1$ in~(\ref{J1_inf}). 
Set the bounds $u_{\max} = 30$, $n_{\max} = 10$ in~(\ref{constraint_on_controls}). The~regularization~(\ref{regularization_in_controls_integral}) is not used in each of the described below three cases. We use GPM-2 (see the iteration formula~(\ref{iteration_formula_GPM2})) with the gradient of the corresponding functional, parameters $\alpha = 3$, $\beta = 0.9$ fixed for the whole number of iterations. For~comparison, GPM-1 (see the iteration formula~(\ref{iteration_formula_GPM1}))
with the same~$\alpha$ is used. With respect to the both terms of the objective~$J_1$, we use the following stopping criterion for GPMs:
\begin{align}
\left((S(\rho^{(k)}(T)) - S(\rho_0))^2 \leq \varepsilon_{{\rm stop},1} \right) \, \& \,
\left(\frac{1}{P}\int\limits_0^T (S(\rho(t)) - S_{\rho_0})^2 dt \leq \varepsilon_{{\rm stop},2} \right).
\label{stopping_criterion_for_problem_with_J1}
\end{align} 
Set $\varepsilon_{{\rm stop},1} = 10^{-6}$ and $\varepsilon_{{\rm stop},2} = 10^{-5}$. 

Consider the following three cases:
(1)~$\varepsilon = 0.1$ and $c^{(0)}=(\sin(2t),0,0)$; (2)~$\varepsilon = 0.1$ and $c^{(0)}=0$; (3)~$\varepsilon = 0$ and $c^{(0)}=0$. For~the GPM computer implementations, we consider piecewise linear interpolation for $u,~n_1,~n_2$ at the uniform time grid with $M = 10^3$ subintervals. 

{\bf Case 1} ($\varepsilon = 0.1$ and $c^{(0)}=(\sin(2t),0,0)$). GPM-2 at the cost of 132 iterations reaches~(\ref{stopping_criterion_for_problem_with_J1}). For~this case, consider the left column of the subfigures in Figure~\ref{fig2}. We see that all the computed controls $u,~n_1,~n_2$ are non-zero here. We see that the graphs of $S(\rho(t))$ (blue solid), degree of purity $P(\rho(t)) = {\rm Tr}\rho^2(t)$, and the Hilbert--Schmidt distance $\| \rho(t) - \frac{1}{4}\mathbb{I}_4 \| {= [{\rm Tr}\big( (\rho(t) - \frac{1}{4}\mathbb{I}_4)^2 \big)]^{1/2}}$ vs $t \in [0,T]$ are close to the constants that relate to the idea of the keeping problem. At~the same time, the~graph of $\| \rho(t) - \rho_0 \|$ is far from constant and shows that this (approximate) keeping relates to sufficiently different 
distances between the system states and $\rho_0$ at various time instances. For~comparison, GPM-1 is used for the same~$c^{(0)}$. Let the largest allowed number of iterations be 500 for this method. At~the cost of 500 iterations, GPM-1 does not reach the stopping criterion~(\ref{stopping_criterion_for_problem_with_J1}), but~the terminal part of $J_1$ is near $3 \times 10^7$ (rather less than $\varepsilon_{{\rm stop},1} = 10^{-6}$) and $\frac{1}{P}\int\limits_0^T (S(\rho(t)) - S_{\rho_0})^2 dt \approx 0.0005$. Thus, both GPM-1 and GPM-2 work good here, but~GPM-2 reaches the criterion at the cost of~132~iterations.  

{\bf Case 2} ($\varepsilon = 0.1$ and $c^{(0)}=(0,0,0)$). Only the initial guess is different, i.e.,~we use the same values~(\ref{values_of_system_parameters}), etc., the~same other settings in GPM-2. At~the cost of 253 iterations, GPM-2 reaches~(\ref{stopping_criterion_for_problem_with_J1}). The~resulting control~$c$ contains the control $u=0$, while both the obtained controls~$n_1,~n_2$ are non-trivial. The~middle column of the subfigures in Figure~\ref{fig2} shows the obtained results. Thus, in~this keeping problem, it is sufficient to adjust only $n_1,~n_2$ under $u=0$. Moreover, note that for $c^{(0)} = 0$, its component $u^{(0)}=0$ is singular in the sense that the corresponding switching function $\mathcal{K}^u(\chi^{(0)}(t),\rho^{(0)}(t)) \equiv 0$ at the whole~$[0,T]$.

{\bf Case 3} ($\varepsilon = 0.1$ and $c^{(0)}=(0,0,0)$). In contrast to the previous case, here we do not take into account the Lamb shift and the dissipator. The~right column of the subfigures in Figure~\ref{fig2} shows that, in this case, the system dynamics achieve the goal of keeping $S(\rho(t))$ at the level~$S(\rho_0)$ at the whole~$[0,T=5]$.  

\subsubsection{Using the Problem~(\ref{J2_inf}) and Genetic Algorithm}

\begin{figure}[h!]
\centering 
\includegraphics[width=\linewidth]{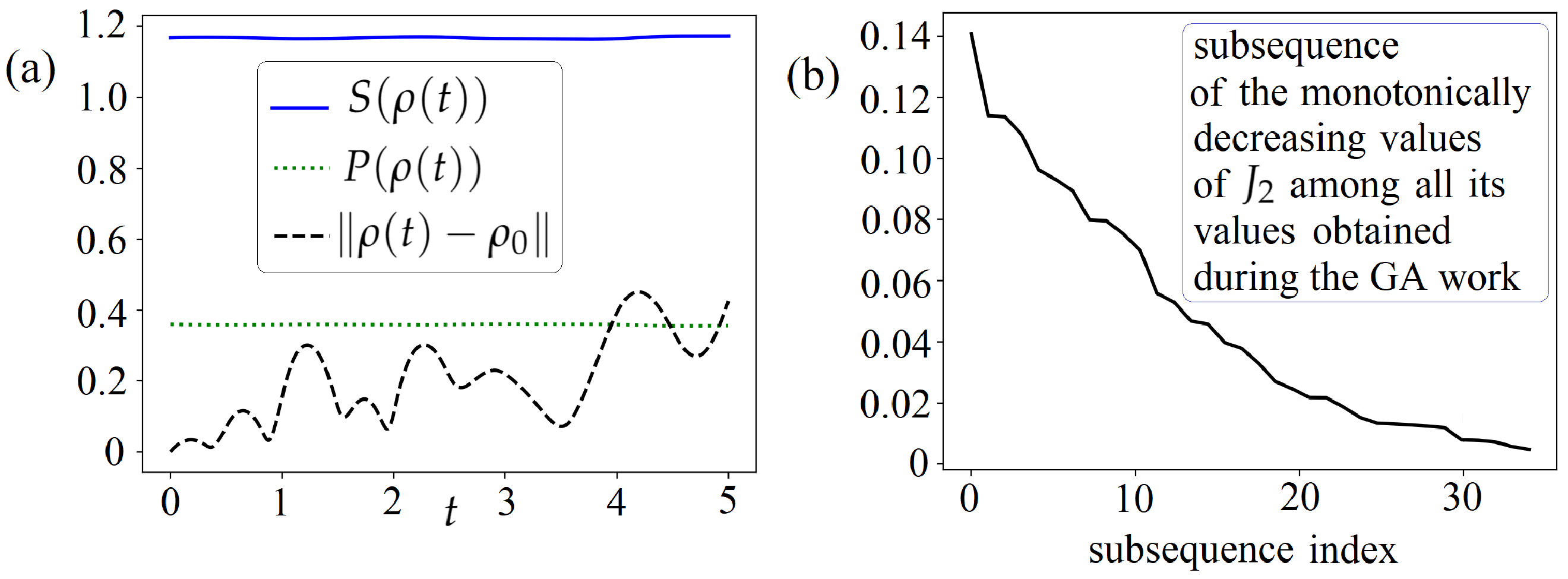}  
\caption{For the problem of keeping the invariant $S(\rho(t)) \equiv S(\rho_0)$ at the whole $[0,T=5]$. Considering piecewise linear controls (with $M=10$ subintervals) relates to the GA finite-dimensional optimization. At~the resulting controls computed with some GA~trial: (\textbf{a})~$S(\rho(t))$, $P(\rho(t))$, and~$\| \rho(t) - \rho_0\|$ vs $t \in [0,T=5]$; (\textbf{b})~the subsequence of the monotonically decreasing values of $J_2$ among all its values computed during the GA~work.\label{fig3}} 
\end{figure}

Further, the~keeping problem is considered as minimizing the objective~$J_2$ in the class of piecewise linear controls via the GA. Here, the class of piecewise linear controls $u,~n_1,~n_2$ is defined at the uniform grid introduced at $[0,T=5]$ with only $M=10$ subintervals (compare with $M=10^3$ used for interpolation of controls in the GPM computer realization). Thus, here, we consider $3(M+1) = 33$ control parameters. Consider $u_{\max} = n_{\max} = 4$ and use the regularization~(\ref{obj_functions_with_regularization2}) with $\gamma_u = 0$, $\gamma_n = 0.01$, $\delta_{n_1} = \delta_{n_2} = 1$. For~GA,  we set the allowed number of iterations to~350. Figure~\ref{fig3} shows the results obtained due to some GA trial that started from an~automatically generated initial point. In~this case, we obtain $J_2 = q_2 \approx 0.005$, satisfying the regularization requirements for incoherent controls in~(\ref{obj_functions_with_regularization2}) with the largest allowed jumps $\delta_{n_1} = \delta_{n_2} = 1$. All~the resulting controls $u,~n_1,~n_2$ are non-trivial~here. 

\subsection{The Problem of Steering the von Neumann Entropy to a~Predefined~Value}  

Consider the steering problem as only the terminal problem, i.e.,~we use the objective~$J_3$ and~(\ref{J3_inf}). As~with objective $J_1$, we also consider the system with the values in Equation~(\ref{values_of_system_parameters}), setting bounds $u_{\max} = 30$, $n_{\max} = 10$. We set the initial state $\rho_0 = {\rm diag}\big(0, \frac{1}{2}, 0, \frac{1}{2} \big)$ with $S(\rho_0) = \log 2 \approx 0.7$ and the target value $S_{\rm tar} = 0.4$. Set $T = 40$. With~respect to the regularization~(\ref{regularization_in_controls_integral}), we consider two cases: with and without this regularization. GPM-2 is used with $\alpha=3$ and $\beta=0.9$. Piecewise linear interpolation for controls is used with \mbox{$M=10^3$} equal subintervals. We take $c^{(0)} = 0.5$. The~stopping criterion is \mbox{$J_3(c^{(k)}) \leq \varepsilon_{\rm stop} = 10^{-6}$}. 

\begin{figure}[h!]
\centering 
\includegraphics[width=\linewidth]{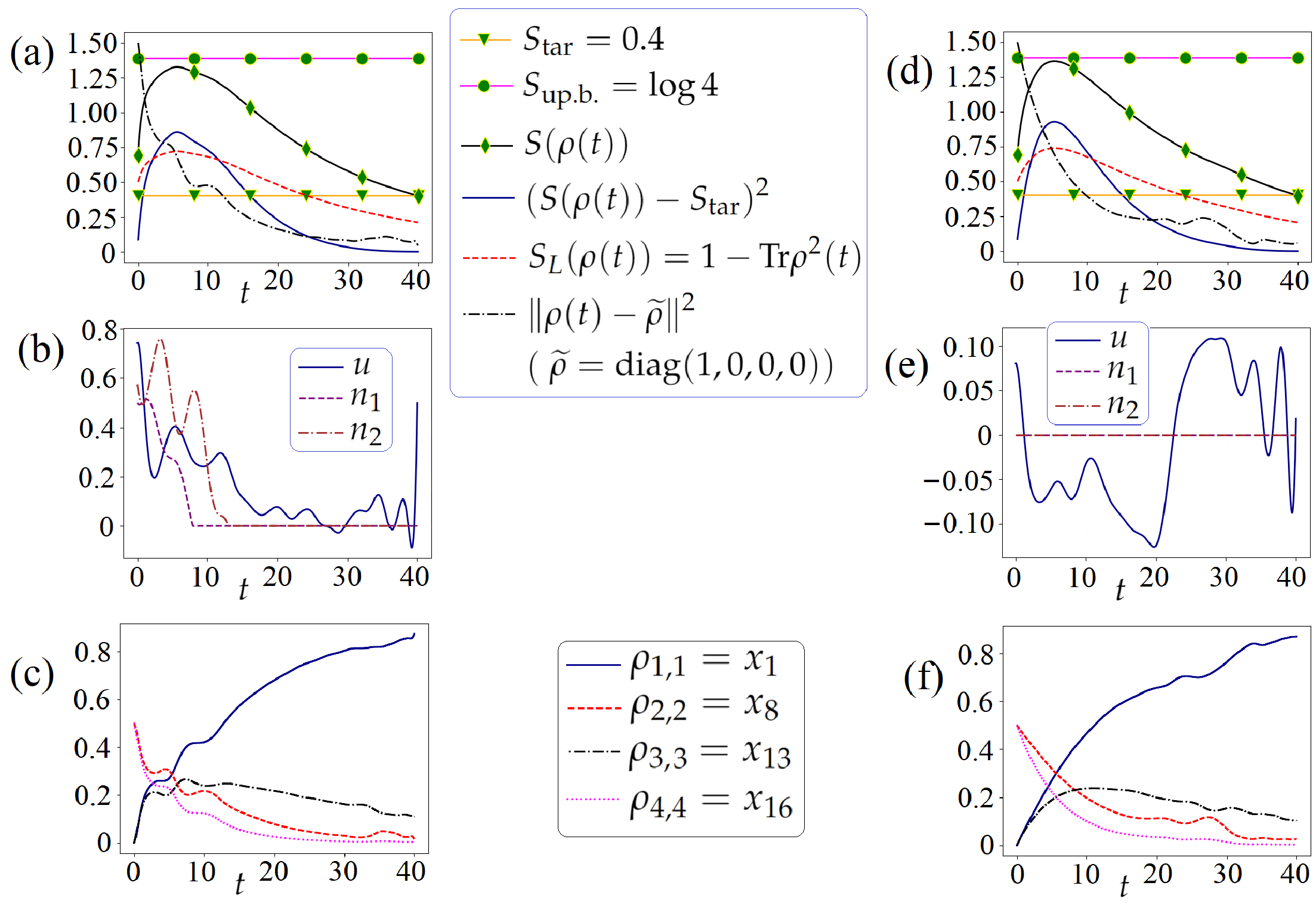}  
\caption{For the problem of steering the von Neumann entropy to the predefined value~$S_{\rm tar}=0.4$ from the initial value $S(\rho_0) \approx 0.7$: without (see the subfigures~(\textbf{a}--\textbf{c})) and with (see the subfigures~(\textbf{d}--\textbf{f})) the regularization~(\ref{regularization_in_controls_integral}). Here, $S_{\rm up.b.} = \log 4$ is {the von Neumann entropy upper bound}, $S_L(\rho(t)) = 1 - P(\rho(t))$ is the linear~entropy.\label{fig4}}  
\end{figure}

{\bf Case 1:} Without the regularization~(\ref{regularization_in_controls_integral}). GPM-2, at the cost of 42 iterations, meets the stopping criterion. The obtained results are shown in Figure~\ref{fig4}a,b,c. We see that all the resulting controls are~non-trivial. 

{\bf Case 2:} With the regularization~(\ref{regularization_in_controls_integral}). Set $\gamma_u = \gamma_n = 10^{-3}$. GPM-2, at the cost of 34 iterations, meets the stopping criterion. The obtained results are shown in Figure~\ref{fig4}d,e,f. We see that only coherent control is computed as non-trivial. Thus, for~the considered steering problem, it is sufficient to adjust only non-trivial coherent~control.

\subsection{The Steering Problem for the von Neumann Entropy under the Pointwise Constraint for This~Entropy} 

In view of the graphs of $S(\rho(t))$ vs $t$ in Figure~\ref{fig4}{a,d}, we introduce and try to satisfy the pointwise constraint $S(\rho(t)) \leq \overline{S} = 1$, $t \in [0, T=40]$, in~addition to the requirement to reach the value $S_{\rm tar} = 0.4$.

Consider both the problems (\ref{J4_inf}) and (\ref{J5_inf}) and, correspondingly, GPM and~GA.

\subsubsection{Using the Problem~(\ref{J4_inf}) and GPM}

\begin{figure}[h!]
\centering 
\includegraphics[width=\linewidth]{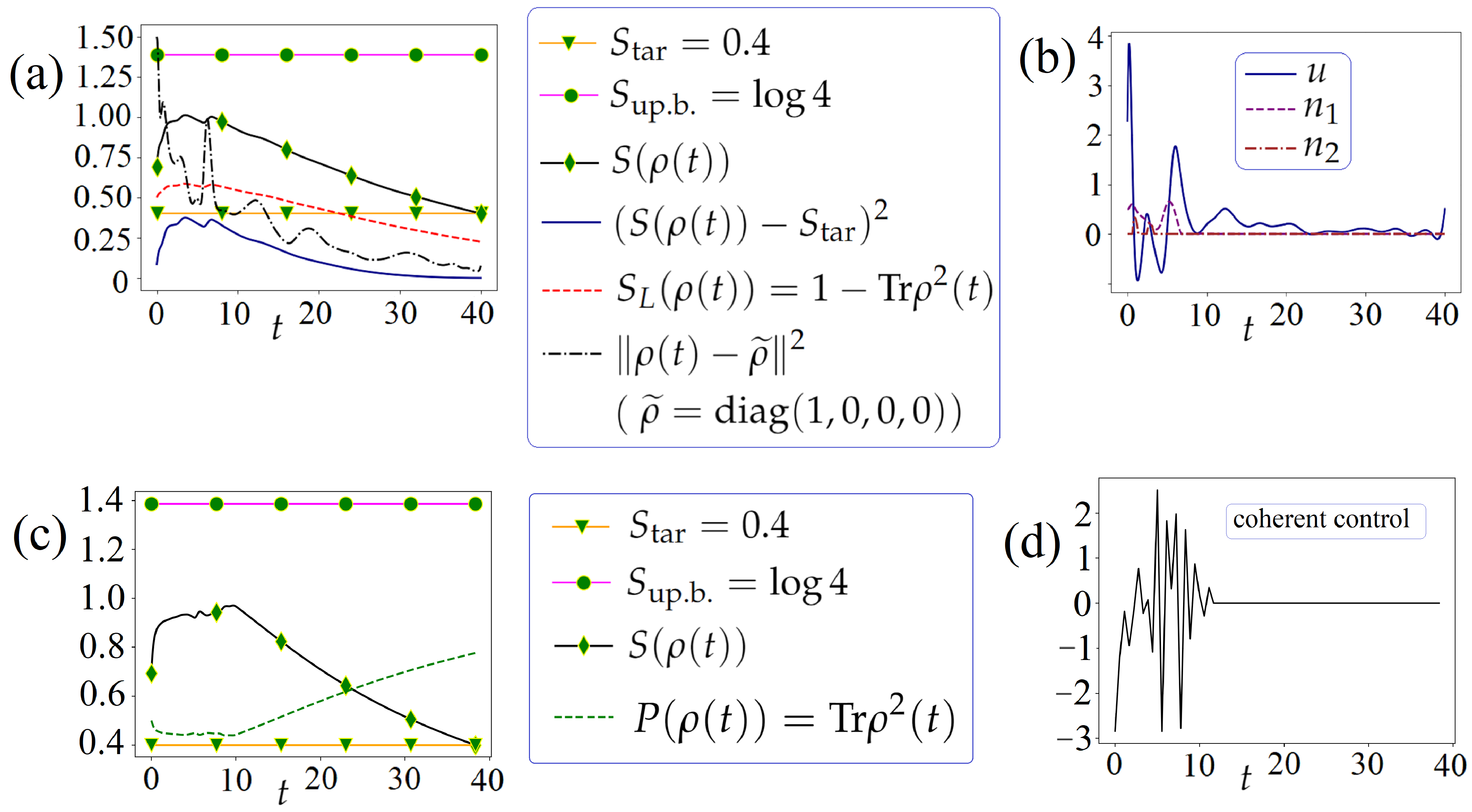}  
\caption{For the problem of steering, the von Neumann entropy to the predefined value~$S_{\rm tar}=0.4$ from the initial value $S(\rho_0) \approx 0.7$ under the state constraint $S(\rho(t)) \leq \overline{S}=1$: (1)~with respect to the problem~(\ref{J4_inf}) (without the regularization~(\ref{regularization_in_controls_integral})) and using GPM-2 (subfigures~(\textbf{a},\textbf{b})); (2)~with respect to the problem~(\ref{J5_inf}) (with the described in the main text special class of controls) and the regularized objective (\ref{obj_functions_with_regularization1}) (with $\gamma_u = 0.1$, $\gamma_n = 0$) and using the GA (subfigures~(\textbf{c},\textbf{d})). We see that, for~approximate steering, it is appropriate to adjust only coherent control under the zero incoherent~controls.\label{fig5}} 
\end{figure}

Consider the objective $J_4$ and the problem~(\ref{J4_inf}). With respect to  both terms of the objective~$J_4$, we use the following stopping criterion for the~GPMs:
\begin{align}
\left((S(\rho^{(k)}(T)) - S_{\rm tar})^2 \leq \varepsilon_{{\rm stop},1} \right) \, \& \,
\left(\frac{1}{P} \int\limits_0^T (\max\{S(\rho^{(k)}(t)) - \overline{S}, 0 \})^2 dt \leq \varepsilon_{{\rm stop},2} \right).
\label{stopping_criterion_for_problem_with_J4}
\end{align} 
Set $\varepsilon_{{\rm stop},1} = 10^{-6}$ and $\varepsilon_{{\rm stop},2} = 10^{-3}$. {We take t}he penalty coefficient~$P = 0.05$ in~$J_4$. The~regularization~(\ref{regularization_in_controls_integral}) is not used here. We set the bounds $u_{\max} = 30$, $n_{\max} = 10$ in~(\ref{constraint_on_controls}). GPM-2 with $\alpha=3$, $\beta=0.9$ at the cost of 39 iterations provides reaching~(\ref{stopping_criterion_for_problem_with_J4}). The~results are shown in Figure~\ref{fig5}{a,b}. 

\unskip

\subsubsection{Using the Problem~(\ref{J5_inf}) and {Genetic Algorithm}}

Consider the problem of steering the von Neumann entropy under the pointwise constraint on $S(\rho(t))$ as minimizing $J_5$. Here, taking into account 
the structure of the resulting controls obtained via GPM-2, and shown in Figure~\ref{fig5}b, we construct the following special class of piecewise linear controls. Let both incoherent controls be zero throughout the interval $[0,T=40]$, while coherent control is zero at $(0.3 T, T]$, and is a~piecewise linear function at $[0, 0.3T]$, which is determined at the uniform grid with $M=20$ subintervals taken at $[0, 0.3T]$. Consider the bound~$u_{\max}=4$ and penalty factor $P=0.5$. In~this optimization problem, $T$ is not fixed and is considered as a~control parameter varied at the range $[T_1,T_2]=[38, 40]$. Thus, here the objective function $g_5$ depends on $M+1 = 21$ control parameters, which determine coherent control, and~$T$. Moreover, the~regularization in the control parameters according to~(\ref{obj_functions_with_regularization1}) is used with $\gamma_u = 0.1$, $\gamma_n = 0$. The~upper bound for the number of iterations of the GA is set at 200. The~results of certain GA~trials are shown in Figure~\ref{fig5}c,d. The~resulting value $|S(\rho(T)) - S_{\rm tar}| \approx 6 \times 10^{-5}$ and the computed pointwise max-max term in~$J_5$ is zero. Thus, we see that, for~approximate steering, it is appropriate to adjust only coherent control under the zero incoherent controls~here. 

\section{Conclusions}
\label{Section_Conclusion}

In this article, we consider the general problem of controlling the von~Neumann entropy of quantum systems either at some final time or over some time interval. The~example of the two-qubit system is considered in detail with  the following control goals: (1)~minimizing or maximizing the final entropy $S(\rho(T))$; (2)~steering $S(\rho(T))$ to a~given target value; (3)~steering $S(\rho(T))$ to a~target value and satisfying the pointwise state constraint $S(\rho(t)) \leq \overline{S}$ 
for a~given~$\overline{S}$; (4)~keeping $S(\rho(t))$ constant at a~given time interval. Under~the Markovian two-qubit dynamics determined by a GKSL-type  
master equation with coherent and incoherent controls: (1)~for the differentiable cases and piecewise continuous controls, one- and two-step gradient projection methods have been adapted by deriving the corresponding adjoint systems and gradients for the objective functionals; (2)~for the non-differentiable cases and piecewise linear controls, a~finite-dimensional optimization with the genetic algorithm has been performed. The numerical experiments conducted with these optimization tools demonstrate their appropriateness for the problems considered and enable the identification of various structures in the resulting controls. A~more detailed analysis of the entropy involving objective functionals, taking into account the Hilbert--Schmidt distances and the reduced density matrices~(\ref{reduced_density_matrices_for_qubits_vs_t}), is an open direction for future research.

\section*{Funding}  
This work was performed at the Steklov International Mathematical Center and supported by the Ministry of Science and Higher Education of the Russian Federation (agreement no.~075-15-2022-265).

\end{document}